\documentclass{aa}  

\usepackage{subcaption}
\usepackage{graphicx}
\usepackage{txfonts}
\usepackage{amsmath}
\usepackage{caption}
\usepackage{subcaption}

\usepackage{lscape}
\usepackage{tocloft}
\usepackage{natbib}
\usepackage{hyperref}
\graphicspath{{./}{}}

\begin{document}

\title{Projection factor and radii of Type II Cepheids \thanks{Tables 1-3 are only available in electronic form at the CDS via anonymous ftp to cdsarc.u-strasbg.fr (130.79.128.5) or via http://cdsweb.u-strasbg.fr/cgi-bin/qcat?J/A+A/.}}
\subtitle{BL Her stars}
\titlerunning{Projection factors and radii of BL Her stars}
\authorrunning{Wielg\'orski, Pietrzyński, Gieren et al.}

\author{P. Wielg\'orski
\inst{1}\thanks{\email{pwielgor@camk.edu.pl}}
\and
G. Pietrzy\'nski
\inst{1,2}
\and
W. Gieren
\inst{2}
\and
B. Zgirski
\inst{2}
\and
M. G\'orski
\inst{1}
\and
J. Storm
\inst{3}
\and
N. Nardetto
\inst{4}
\and
P. Kervella
\inst{5}
\and
G. Bras
\inst{5}
\and
G. Hajdu
\inst{1}
\and
V. Hocd\'e
\inst{1}
\and
B. Pilecki
\inst{1}
\and
W. Narloch
\inst{1}
\and
P. Karczmarek
\inst{2}
\and
W. Pych
\inst{1}
\and
R. Chini
\inst{1,6,7}
\and
K. Hodapp
\inst{8}
}

\institute{Nicolaus Copernicus Astronomical Center, Polish Academy of Sciences, Bartycka 18, 00-716 Warszawa, Poland
\and
Universidad de Concepci\'on, Departamento de Astronomia, Casilla 160-C, Concepci\'on, Chile
\and
Leibniz-Institut f\"ur Astrophysik Potsdam (AIP), An der Sternwarte 16, 14482 Potsdam, Germany
\and
Université Côte d'Azur, Observatoire de la Côte d'Azur, CNRS, Laboratoire Lagrange, France
\and
LESIA, Observatoire de Paris, Universit\'e PSL, CNRS, Sorbonne Universit\'e, Universit\'e Paris Cit\'e, 5 place Jules Janssen, 92195 Meudon,France
\and
Astronomisches Institut, Ruhr-Universit\"at Bochum, Universit\"atsstrasse 150, D-44801 Bochum, Germany
\and
Instituto de Astronom\'{i}a, Universidad Cat\'{o}lica del Norte, Avenida Angamos 0610, Antofagasta, Chile
\and
University of Hawaii, Institute for Astronomy, 640 N. Aohoku Place, Hilo, HI 96720, USA}

\date{Received ..., 2024; accepted ...., 2024}

\abstract
    {Type II Cepheids are old pulsating stars that can be used to trace the distribution of an old stellar population and to measure distances to globular clusters and galaxies within several megaparsecs, and by extension, they can improve our understanding of the cosmic distance scale. One method that can be used to measure the distances of Type II Cepheids relies on period-luminosity relations, which are quite widely explored in the literature. The semi-geometrical Baade-Wesselink technique is another method that allows distances of radially pulsating stars, such as Type II Cepheids, to be measured if the so-called projection factor is known. However, the literature concerning this parameter for Type II Cepheids is limited to just a few pioneering works.}
    {In determining projection factors for eight nearby short-period Type II Cepheids, also known as BL Her type stars, we aim to calibrate the Baade-Wesselink method for measuring distances for this class of stars.}
    {Using the surface brightness-colour relation version of the Baade-Wesselink technique, we determined the projection factors and radii of eight nearby BL Her type stars. We adopted accurate distances of target stars from $Gaia$ Data Release 3. Time series photometry in the $V$ and $K_{\mathrm{S}}$ bands have been collected with two telescopes located at the $Rolf$ $Chini$ Cerro Murphy Observatory (former Cerro Armazones Observatory), while spectroscopic data have been obtained within dedicated programmes with instruments hosted by the European Southern Observatory.}
    {The measured projection factors for the stars with good quality data are in the range between 1.21 and 1.36. The typical uncertainty of projection factors is 0.1. The mean value is 1.330$\pm$0.058, which gives the uncertainty of $\sim$4\%. The main sources of uncertainty on the $p$-factors are statistical errors of the Baade-Wesselink fit (related to the dispersion and coverage of light and radial velocity curves) and parallax. In the case of radii, the biggest contribution to the error budget comes from the $K_{\mathrm{S}}$ band photometry's systematic uncertainty and parallax. The determined radii allowed us to construct the period-radius relation for BL Her stars. Our period-radius relation is in good agreement with the previous empirical calibration, while two theoretical calibrations found in the literature agree with our relation within 2$\sigma$. We also confirm that BL Her and RR Lyr stars obey an apparent common period-radius relation.}
    {}

\maketitle

\section{Introduction} \label{sec:intro}
Radially pulsating stars, such as Cepheids (classical, Type II, or anomalous) and RR Lyrae, serve as very important distance indicators. What makes them very special is that there are two completely independent methods of determining distances to these stars that take advantage of their pulsational nature. The respective period-luminosity relations of pulsating stars  \citep[Leavitt Law;][]{1908AnHar..60...87L} make them widely usable as very precise standard candles \citep[e.g.][]{2005ApJ...628..695G,2017MNRAS.472..808R,2021ApJ...908L...6R,2016ApJ...832..210B,2020JApA...41...23B,2023ApJ...951..118B,2024AA...685A..41S}. In particular, the period-luminosity relation for classical Cepheids is used in the most precise distance ladder to determine the $Hubble$ $Constant$ \citep{2023ApJ...956L..18R}. On the other hand, the semi-geometric Baade-Wesselink (BW) method \citep[it is also sometimes called the Baade-Becker-Wesselink or parallax-of-pulsation method;][]{1926AN....228..359B,1940ZA.....19..249B,1947BAN....10..252W} can be applied to measure the distance to the radially pulsating star by comparing variations of the physical (inferred from integration of the radial velocity curve) and apparent angular size (obtained from photometry or interferometry). Different versions of the BW method have been applied by many authors to obtain distances and sizes of pulsating stars \citep[e.g.][]{1977MNRAS.178..231B,1986AA...159..261B,1998ApJ...496...17G,2011AstBu..66...47R,2013AA...550A..70G,2014MNRAS.437..906R,2015A&A...584A..80M}. This method requires prior knowledge of the so-called projection factor ($p$-factor), which translates the radial velocity measured from spectra into the velocity of the expanding or contracting atmosphere of the star (pulsational velocity). The value of the $p$-factor is the result of the projection of the pulsational velocity from a given part of the visible stellar disc on the line of sight. Due to limb darkening, velocity gradients in the atmosphere, and other factors, the value of the $p$-factor is difficult to model \citep[see, e.g.,][and references therein]{2011AA...534L..16N,2014IAUS..301..145N}. An accurate calibration of this parameter is crucial to achieve high accuracy on distances measured with the BW method. 

Empirical determination of the projection factor requires prior knowledge of the distances of a sample of stars of a given class. While parallaxes of a number of classical Cepheids and RR Lyrae stars were measured by $Hipparcos$ and the $Hubble$ Space Telescope, the Type II Cepheid distance scale was based on parallaxes of just two stars before $Gaia$ \citep{2009MNRAS.397..933M,2011AJ....142..187B}. The $Gaia$ space mission \citep{2016AA...595A...1G} is a game-changer, and the last data release \citep[Data Release 3, DR3, ][]{2021AA...649A...1G,2021AA...649A...2L} contains parallaxes of several dozens of Type II Cepheids within 5kpc. In \citet{2022ApJ...927...89W}, we used $Gaia$ DR3 parallaxes of 21 Galactic Type II Cepheids to determine period-luminosity relations with an accuracy of $\sim$5\%.

Different versions of the BW analysis have been used to determine radii of Type II Cepheids assuming the value of the $p$-factor \citep[e.g. 1.36,][]{1977MNRAS.178..231B,1986AA...159..261B,1997AJ....113.1833B}. The literature concerning $p$-factors of Type II Cepheids is limited to just some pioneering works \citep{2008MNRAS.386.2115F,2015A&A...576A..64B,2018ApJ...868...30P}. Two stars analysed in these papers are peculiar W Virginis (pW Vir) stars, which are believed to be members of binary systems, which is a quite strong limitation for using them as precise distance indicators. \citet{2008MNRAS.386.2115F} used the Hipparcos parallax of $\kappa$ Pav in an analysis that is similar to what is presented in this paper and obtained a very low $p$-factor value of 0.93$\pm$0.11. \citet{2015A&A...576A..64B} analysed $\kappa$ Pav with the Spectro-Photo-Interferometry of Pulsating Stars code \citep[\texttt{SPIPS})][]{2015A&A...584A..80M}, which utilises multiband photometric and interferometric datasets and models of stellar atmospheres to determine the angular size of the star, and they obtained p=1.26$\pm$0.07. We note that $\kappa$ Pav is believed to belong to the pW Vir class \citep{2009MNRAS.397..933M,2022ApJ...927...89W}; however, no signature of a companion has been found so far. On the other hand, \citet{2018ApJ...868...30P} performed a very detailed analysis of pW Vir found by the Optical Gravitational Lensing Experiment \citep[$OGLE$,][]{2008AcA....58...69U,2008AcA....58..293S} project in the Large Magellanic Cloud, which is a known eclipsing binary. The value of the projection factor obtained in that study is 1.30$\pm$0.03.

Recent studies of $p$-factors for classical Cepheids \citep{2021AA...656A.102T} and RR Lyrae stars \citep{2024AA...684A.126B} with \texttt{SPIPS} and using accurate parallaxes from the $Gaia$ satellite mission have revealed quite a big scatter ($\sim$10\%) and no clear correlation of $p$-factors with the period. On the other hand, \citet{2017AA...608A..18G} in their analysis of classical Cepheids in the Large Magellanic Cloud with \texttt{SPIPS} obtained a quite significant linear dependence of the $p$-factor on the period. The observed large scatter (not compatible with the statistical errors of the measured $p$-factors) of $p$-factors could be the result of the non-uniform datasets used in the analysis (different instruments and epochs of observations); cycle-to-cycle variations of the radial velocity amplitude \citep{2016MNRAS.455.4231A,2016MNRAS.463.1707A}; the gradient of the velocity in the stellar atmosphere \citep{2017AA...597A..73N}; or even envelopes composed of dust or ionised gas surrounding these stars \citep{2009AA...498..425K,2020AA...633A..47H,2023AA...671A..14N,2024AA...683A.233H}. This may also suggest that using a single parameter to translate the observed radial velocity into the pulsational velocity of the stellar atmosphere is too great of a simplification, but we believe that there is still some space for improvements both in the data and methodology. Comparing projection factors for different types of pulsating stars with different physical properties can also be very helpful in improving our understanding of stellar atmospheres.

In this study, we present the very first determination of the $p$-factor for Type II Cepheids of the BL Her subclass using the surface brightness-colour relation (SBCR) version of the BW method \citep[][]{1976MNRAS.174..489B,1997AA...320..799F}. Our sample consists of eight stars, but it will be expanded in the future, which should allow us to check the intrinsic scatter of the $p$-factor and its dependence on other parameters. This work is part of a series of papers on the Araucaria project \citep{2023arXiv230517247A} devoted to the calibration of primary distance indicators using nearby stars and is based on observations made at the $Rolf$ $Chini$ Cerro Murphy Observatory \citep[OCM; formerly the Cerro Armazones Observatory;][]{2016SPIE.9911E..2MR} and $Gaia$ parallaxes. Thanks to the use of uniform datasets, as described in Section \ref{sec:data}, we minimised the influence of systematic errors on our results. In Section \ref{sec:method}, we present our methodology, and Section \ref{sec:results} presents our results and discussion. We summarise our work in Section \ref{sec:summ}.

\section{Data} \label{sec:data}

\subsection{Optical and near-infrared photometry, reddening, and periods}

The photometric data were collected between March 2017 and March 2020 with two telescopes located at OCM. Near-infrared time series data in the $J$, $H$, and $K_{\mathrm{S}}$ passbands, collected with the IRIS camera \citep{2010SPIE.7735E..1AH} installed on the 0.8-metre telescope, have been published in \citet{2022ApJ...927...89W}. In this work, we used the $K_{\mathrm{S}}$ band data only. We supplemented our $K_{\mathrm{S}}$ band measurements for SW Tau with data from \citet{2008MNRAS.386.2115F}, collected with a photometer installed on the 0.75-metre telescope at the South African Astronomical Observatory (SAAO). The IRIS photometry is in the Two Micron All Sky Survey \citep[2MASS, ][]{2006AJ....131.1163S} system, and we transformed the SAAO photometry to the 2MASS system using formulae from \citet{2007MNRAS.380.1433K}. The systematic uncertainty of the $K_{\mathrm{S}}$ band light curves was estimated in \citet{2022ApJ...927...89W} to 0.02mag.

Optical photometry in the Johnson $B$ and $V$ passbands were obtained with the $VYSOS16$ 0.4-metre telescope located at OCM \citep{2013AN....334.1115R} with a pixel scale of 0.77 arcsec/px and a 41.2'x25.5' field of view. Nine dithered images were corrected for dark current and flatfielded with the standard \texttt{IRAF} routines \citep{1993ASPC...52..173T} and then matched and stacked into one final frame using \texttt{Sextractor} \citep{1996AAS..117..393B}; \texttt{SCAMP} \citep{2006ASPC..351..112B}; and \texttt{SWARP} \citep{2010ascl.soft10068B}. Aperture photometry was made with a dedicated pipeline based on the \texttt{Astropy} Python library \citep{2018AJ....156..123A}, and \texttt{DAOPHOT} \citep{1987PASP...99..191S} was transformed to the standard Johnson-Cousins system using secondary standards from the synthetic all sky catalogue described in \citet{2023AA...674A..33G}, which was created based on $Gaia$ low-resolution spectra. The synthetic catalogue was standardised with \citet{1992AJ....104..340L} standard fields, and the authors estimated its accuracy with secondary standards from the `Stetson collection' \citep[for details see section 3.2 in ][]{2023AA...674A..33G} to be at the level of $\sim$0.02 mag. We adopted this value as the systematic uncertainty of our transformations. We selected as secondary standards three to ten stars with a brightness comparable to the target star and a wide range of colours. The $(B-V)$ colour term of our transformations to the standard system is $-$0.04 for the $B$ band and zero for the $V$ band. The $B$ and $V$ band photometric data are presented in Table \ref{tab:vysos16}.

We adopted periods published in \citet{2022ApJ...927...89W} and presented in Table \ref{tab:data}. In the case of SW Tau, the period given in the table is for $HJD=2435500.0$, and we applied the period change of $-1.15\times10^{-6}$s/yr from \citet{2017AJ....153..102D} to improve the quality of curves, in particular to phase the $K_{\mathrm{S}}$ band measurements from IRIS and \citet{2008MNRAS.386.2115F} and by extension to minimise the spread of the final fit.

\begin{figure}[h]
            \centering
            \includegraphics[width=0.5\textwidth]{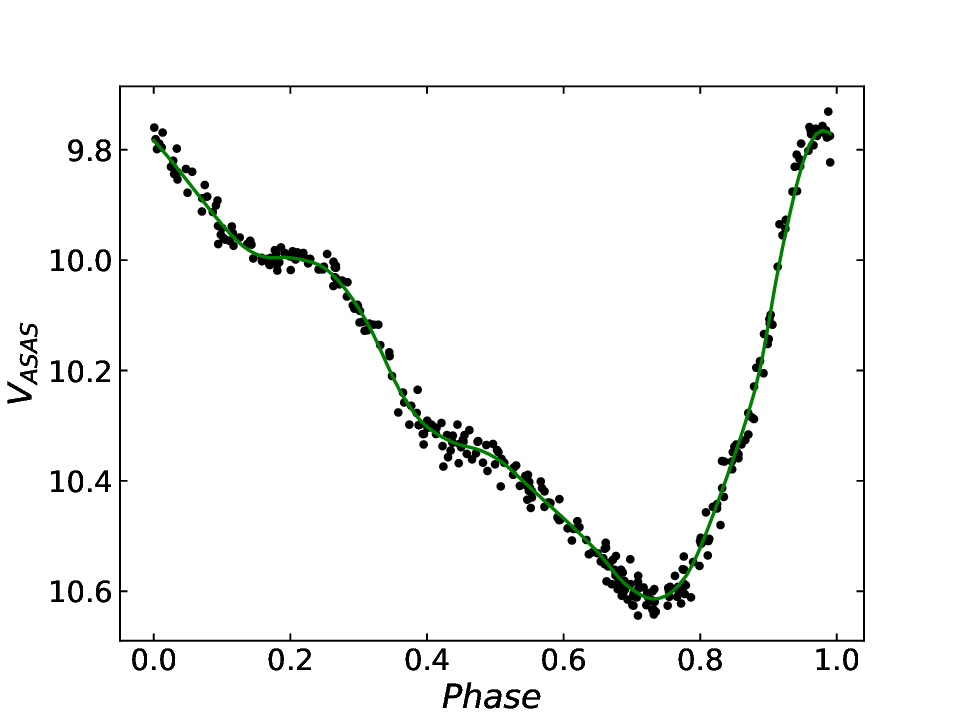}
            \caption{$ASAS$ photometry used to create the $V$ band light curve template for BL Her star. The green line is the Akima spline fit.\label{fig:blher_templ}}
            \end{figure}

Our light curves are quite well sampled, but in some cases we decided to fit templates to our measurements in the $V$ band. Templates were created using photometry of a given star in the $V$ band from the All Sky Automated Survey \citep[ASAS;][]{2004AN....325..553P} or from the All Sky Automated Survey for Supernovae \citep[ASASSN;][]{2014AAS...22323603S,2018MNRAS.477.3145J}, depending on the quality of the photometry from a given source. The ASAS or ASASSN photometry was phased with a period given in Table \ref{tab:data} and approximated with Akima splines \citep{10.1145/321607}, implemented in the \texttt{SciPy} Python library \citep{2020SciPy-NMeth}. The template light curve was then fit (amplitude, phase shift, and mean magnitude) to our measurements using $\chi ^2$ minimisation. In Figure \ref{fig:blher_templ}, we show the exemplary photometry used to create the template.

The photometry was corrected for the interstellar extinction using values of the colour excess $E(B-V)$ published in \citet{2022ApJ...927...89W}, which are based on \citet{2011ApJ...737..103S} reddening maps and the dust model in the Milky Way from \citet{2001ApJ...556..181D}. To calculate the total extinction in $V$ and $K_{\mathrm{S}}$ bands, we used \citet{1989ApJ...345..245C} and \citet{1994ApJ...422..158O} reddening law, assuming $R_V$=3.1. The adopted values of $E(B-V)$ are also presented in Table \ref{tab:data}. The statistical uncertainty given in \citet{2011ApJ...737..103S} is below 0.01 mag, but we assumed the uncertainty of the $E(B-V)$ values to be at the level of 0.02mag.

\begin{table} 
    \caption{New optical photometry of the Type II Cepheid sample analysed in this paper. The full version of this table is available as supplementary material in a machine readable format at the CDS.}
    \label{tab:vysos16}
    \centering
    \begin{tabular}{ccccc}
    \hline\hline
   
    $Star$ & $Filter$ & $HJD$ & $m$ & $\sigma _{m}$\\
    & & (days) & (mag) & (mag) \\
    \hline
    BL Her &	V	& 2458342.531808 & 9.792 & 0.005\\
BL Her	&	V	&	2458344.533328	&	10.386	&	0.005\\
BL Her	&	V	&	2458346.538110	&	9.894	&	0.005\\
BL Her	&	V	&	2458347.533630	&	10.441	&	0.005\\
BL Her	&	V	&	2458348.547888	&	10.470	&	0.005\\
BL Her	&	V	&	2458349.540027	&	10.226	&	0.005\\
BL Her	&	V	&	2458350.536321	&	9.963	&	0.005\\
BL Her	&	V	&	2458351.534698	&	10.227	&	0.005\\
BL Her	&	V	&	2458352.530724	&	10.548	&	0.005\\
BL Her	&	V	&	2458353.525130	&	10.319	&	0.005\\
BL Her	&	V	&	2458355.525918	&	9.837	&	0.005\\
BL Her	&	V	&	2458356.519569	&	10.601	&	0.005\\
BL Her	&	V	&	2458357.494819	&	10.330	&	0.005\\
BL Her	&	V	&	2458358.495066	&	10.001	&	0.005\\
BL Her	&	V	&	2458359.495035	&	9.770	&	0.005\\
BL Her	&	V	&	2458368.551779	&	10.090	&	0.005\\
BL Her	&	B	&	2458342.525882	&	10.045	&	0.009\\
BL Her	&	B	&	2458344.527333	&	10.929	&	0.006\\
BL Her	&	B	&	2458346.532184	&	10.188	&	0.006\\
BL Her	&	B	&	2458347.527763	&	10.979	&	0.005\\
BL Her	&	B	&	2458348.531766	&	11.041	&	0.005\\
    \hline
    \end{tabular}
    \end{table}

    \begin{table*}
        \caption{Periods, parallaxes, parallax zero-point offset, $Gaia$ $G$-band magnitude and $(Bp-Rp)$ colour index, and reddening for the analysed Type II Cepheids.}
        \label{tab:data}
        \centering
        \begin{tabular}{ccccccc}
        \hline\hline
       
        $Name$ & $P$ & $\omega$ & $ZPO$ & $G_{Gaia}$ & $(Bp-Rp)$ & $E(B-V)$\\
         & (days) & (mas) & (mas) & (mag) & (mag) & (mag)\\
        \hline
        VY Pyx & 1.239957 & 3.9495$\pm$0.0186 & -0.0237 & 7.11 & 0.81 & 0.048$\pm$0.02\\
        BL Her & 1.30743 & 0.8469$\pm$0.0179 & 0.0016 & 10.17 & 0.74 & 0.067$\pm$0.02\\
        KZ Cen & 1.52006 & 0.3024$\pm$0.0153& -0.0224 & 12.37 & 0.74 & 0.084$\pm$0.02\\
        SW Tau & 1.58355* & 1.2244$\pm$0.0222 & -0.0103 & 9.62 & 0.98 & 0.252$\pm$0.02\\
        V971 Aql & 1.62453 & 0.4400$\pm$0.0219 & -0.0175 & 11.78 & 0.91 & 0.174$\pm$0.02\\
        V439 Oph & 1.89298 & 0.4753$\pm$0.0163 & -0.0103 & 11.89 & 1.15 & 0.268$\pm$0.02\\    
        RT TrA & 1.94612 & 1.0162$\pm$0.0162 & -0.0021 & 9.68 & 0.92 & 0.112$\pm$0.02\\
        FM Del & 3.95498 & 0.2300$\pm$0.0135 & -0.0174 & 12.27 & 0.92 & 0.087$\pm$0.02\\
        \hline
        \multicolumn{4}{l}{*Period for $HJD$=2435500.0, period change $-$1.15$\times$10$^{-6}$s/yr}
        \end{tabular}
        \end{table*}

\subsection{Spectroscopy}

Radial velocities were measured from spectra collected between 2016 and 2023 within dedicated programmes with four high resolution (40000-100000) spectrographs hosted by the European Southern Observatory (ESO). We used CORALIE, installed on the 1.2m `Swiss' Leonhard Euler telescope at La Silla Observatory \citep{2000AA...354...99Q}; HARPS, installed on the 3.6m telescope at La Silla Observatory \citep{2000SPIE.4008..582P}; FEROS, installed on the 2.2m MPG/ESO telescope at La Silla Observatory \citep{1998SPIE.3355..844K}; and UVES, installed on the 8.2m VLT UT2 (Kueyen) telescope at the Paranal Observatory \citep[][ filler programme]{2000SPIE.4005..121D}. 

In the case of HARPS and UVES spectra, we used the standard reduction pipelines (instrumental calibrations and wavelength solution) offered by ESO and downloaded processed spectra from the ESO archive. The CORALIE and FEROS spectra were reduced with \texttt{CERES} pipeline \citep{2017PASP..129c4002B}, which proved to give good results for these two instruments. If needed, we merged the echelle orders with a custom Python script developed by our team and normalised spectra by modelling continuum with polynomials (after excluding the most prominent lines). As shown by \citet{2008AA...489.1255N} and \citet{2019AA...631A..37B}, the selection of the analysed spectral lines has an impact on the obtained value of the $p$-factor. To homogenise the spectra from all instruments, we restricted the analysis to the wavelength range between 450nm and 650nm. Moreover, we used a mask to exclude telluric lines and problematic regions (such as gaps) present in some spectra.

Radial velocities were measured using the cross-correlation function (CCF) implemented in \texttt{iSpec} software \citep{2014AA...569A.111B} using \citet{2005AA...443..735C} synthetic spectra as templates. We also tried the Broadening Function (BF) technique implemented in \texttt{Ravespan} \citep{2017ApJ...842..110P}, but it gave virtually identical results. The important concern that quite significantly impacts the obtained results is the selection of the function to model the CCF, as the profiles of lines for pulsating stars are asymmetric in some phases. We decided to use the Gaussian profile to be consistent with \citet{2021AA...656A.102T} and \citet{2024AA...684A.126B}, but we also tried a bi-Gaussian profile and first moment as described in \citet{2017AA...597A..73N}, and we discuss the influence of this choice on the results. The typical uncertainty of our radial velocity measurements is 0.2 km/s. We did not observe any obvious systematic shift between the measurements from different instruments. The measured radial velocities are presented in Table \ref{tab:rvs}.

\begin{table*} 
    \caption{New radial velocity measurements of the sample Type II Cepheids analysed in this paper obtained from a cross-correlation function modelled with Gaussian ($V_{r,G}$), centroid ($V_{r,C}$), and bi-Gaussian ($V_{r,B}$). The full version of this table is available as supplementary material in a machine readable format at the CDS.}
    \label{tab:rvs}
    \centering
    \begin{tabular}{ccccccc}
    \hline\hline
$Star$ & $HJD$ & $V_{r,G}$ & $\sigma _{V_r}$ & $V_{r,C}$ & $V_{r,B}$ & $instrument$\\
& (days) & (km/s) & (km/s) &  (km/s) & (km/s) & \\
\hline
BL Her	& 2457901.69869132	& 18.11	& 0.20	& 25.95	& 18.17	& HARPS\\
BL Her	& 2457901.77199208	& -2.07	& 0.13	& -0.98	& -2.10	& HARPS\\
BL Her	& 2457902.68274749	& 20.32	& 0.07	& 18.17	& 20.36	& HARPS\\
BL Her	& 2457903.75918934	& 10.23	& 0.08	& 12.32	& 10.19	& HARPS\\
BL Her	& 2457914.67843204	& 36.17	& 0.13	& 32.12	& 36.22	& HARPS\\
BL Her	& 2457914.76888472	& 20.27	& 0.32	& 26.88	& 20.27	& HARPS\\
BL Her	& 2457915.62648906	& 12.07	& 0.06	& 12.01	& 12.09	& HARPS\\
BL Her	& 2457915.72868948	& 18.09	& 0.06	& 16.44	& 18.14	& HARPS\\
BL Her	& 2457916.67453464	& 8.31	& 0.06	& 8.24	& 8.29	& HARPS\\
BL Her	& 2458233.77712936	& 26.60	& 0.28	& 29.33	& 26.59	& HARPS\\
BL Her	& 2458240.78722272	& 3.05	& 0.05	& -1.86	& 3.11	& CORALIE\\
BL Her	& 2458240.8691601	& 6.04	& 0.05	& 4.90	& 6.02	& CORALIE\\
BL Her	& 2458240.92456825	& 8.57	& 0.04	& 7.41	& 8.58	& CORALIE\\
BL Her	& 2458241.78463846	& -10.76	& 0.16	& -8.45	& -10.81	& CORALIE\\
BL Her	& 2458241.86983985	& -8.34	& 0.13	& -8.51	& -8.30	& CORALIE\\
BL Her	& 2458242.82536589	& 36.15	& 0.07	& 32.66	& 36.17	& CORALIE\\
BL Her	& 2458243.87606979	& 16.88	& 0.05	& 17.38	& 16.92	& CORALIE\\
\hline
\end{tabular}
\end{table*}

\subsection{Distances}

Similar to \citet{2022ApJ...927...89W}, we used parallaxes from $Gaia$ DR3 with the position, magnitude, and colour-dependent zero-point offset calculated with the dedicated Python code described in \citet{2021A&A...649A...4L}. We increased the statistical uncertainties of the parallaxes given in the $Gaia$ catalogue by 10\% as suggested by \citet{2021ApJ...908L...6R} to account for possible excess uncertainty. The values of the parallaxes and zero-point corrections as well as the $Gaia$ $G$-band magnitudes and $(Bp-Rp)$ colour index used to calculate the zero-point corrections are given in Table~\ref{tab:data}.

The renormalised unit weight error (RUWE) and goodness of fit (GOF) parallax quality parameters given in the $Gaia$ catalogue \citep[see Table 2 in ][]{2022ApJ...927...89W} for the analysed stars are well below the limits for good quality parallaxes \citep[$RUWE$<1.4, $GOF$<12.5,][]{2021A&A...649A...2L}. The distances of the stars in our sample are between 250$\,$pc and 4500$\,$pc, with a mean value of 900$\,$pc and typical error of $\sim$2\%.

\section{The IRSB Baade-Wesselink method} \label{sec:method}

\begin{table*}
    
    \caption{Surface brightness-colour relations used to estimate the angular diameters of Type II Cepheids.\label{tab:sbcr}}
  \centering
  \begin{tabular}{c|c|c|c}

  \hline
 \hline
 abbreviation & relation & calibrating sample & source\\
 \hline
  K04a & $F_V$=-0.1336$\times (V-K)_0$+3.9530 & classical Cepheids & \citet{2004AA...428..587K}\\
  &&0.95 $< (V-K_{\mathrm{S}})_0 <$ 2.4 & \\
  &&& \\
  K04b & $F_V$=-0.1377$\times (V-K)_0$+3.9620 & dwarfs and subgiants & \citet{2004AA...426..297K}\\
  &&0.5 $< (V-K_{\mathrm{S}})_0 <$ 3 & \\
  &&&\\
  S21 & $F_V$=-0.1404$\times(V-K)_0$+3.9665  & dwarfs and subgiants & \citet{2021AA...652A..26S}\\
  &&1 $< (V-K_{\mathrm{S}})_0 <$ 3 & \\
  &&&\\
  G21& $F_V$=-0.0031$\times (V-K)_0 ^5$+0.0239$\times (V-K)_0 ^4$ & dwarfs & \citet{2021AA...649A.109G}\\
  &-0.0623$\times (V-K)_0 ^3$+0.0705$\times (V-K)_0 ^2$& 0.5 $< (V-K_{\mathrm{S}})_0 <$ 2.1 &\\
  &-0.1708$\times (V-K)_0$+3.9666 & &\\
    \hline
  \end{tabular}
  \end{table*}

\begin{figure}[h]
    \centering
    \includegraphics[width=0.5\textwidth]{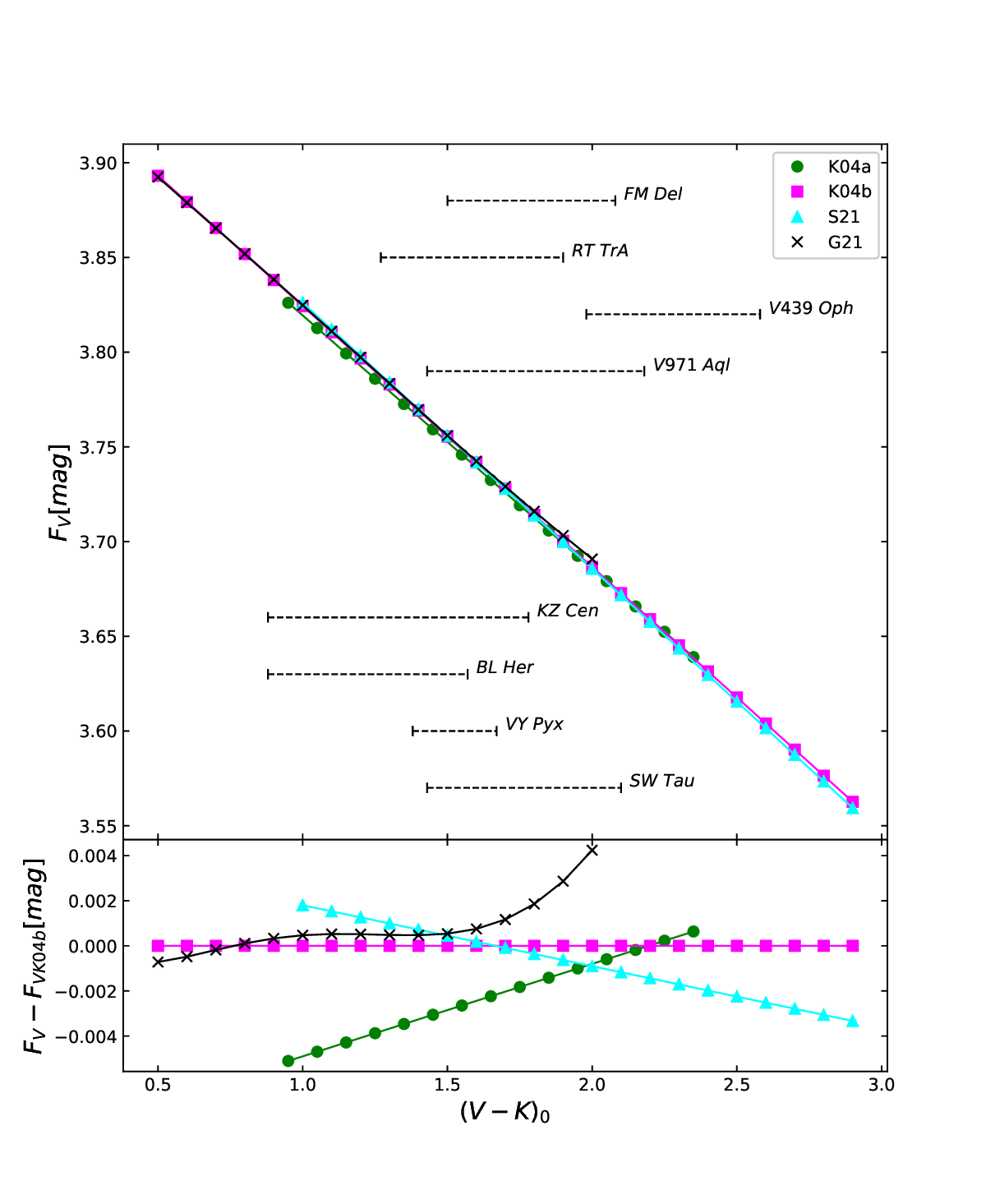}
    \caption{Surface brightness-colour relation used to determine the angular diameters of the analysed stars. The dashed sections denote the colour index range covered by each analysed star during the pulsation cycle.\label{fig:sbcr}}
    \end{figure}

Our implementation of the BW method is similar to the approach presented in \citet{2011AA...534A..94S} and \citet{2018AA...620A..99G}. It relies on the angular diameters determined from the SBCR \citep{1976MNRAS.174..489B} calibrated for the $(V-K_{\mathrm{s}})$ colour index \citep[IRSB technique,][]{1997AA...320..799F}. The angular diameters determined this way very weakly depend on reddening, metallicity, or surface gravity \citep{1997AA...320..799F,2004AA...428..587K,2018AA...620A..99G,2019Natur.567..200P,2001AJ....121.3089T,2022AA...662A.120S}. The surface brightness $F_V$ is defined as follows: 

\begin{equation}\label{eq:sb}
    F_V(\phi)=4.2207 - 0.1V_0 (\phi)- 0.5 \log \theta (\phi),
\end{equation}
where $V_0$ is the unreddened $V$ band magnitude and $\theta$ is the angular diameter. Calibrations of the SBCR have been made for different types of stars by, for example, \citet{1994AJ....108.1421W,1997AA...320..799F,1998AA...339..858D,2004AA...426..297K,2004AA...428..587K,2014AA...570A.104C,2019Natur.567..200P,2021AA...649A.109G,2021AA...652A..26S} from interferometric measurements of angular diameters or eclipsing binaries with known distances. There is no specific relation calibrated for Type II Cepheids (which are giants of F-G spectral type) available, so we decided to use the relations from \citet[][hereafter K04a]{2004AA...428..587K}; \citet[][K04b]{2004AA...426..297K}; \citet[][S21]{2021AA...652A..26S}; and \citet[][G21]{2021AA...649A.109G}, which are presented in Table \ref{tab:sbcr}. We decided to use these relations because they are independent and very precise (dispersion $\sim$1-2\%) and cover the colour range of the analysed Type II Cepheids. These relations are calibrated for classical Cepheids, dwarfs, and subgiants, and such an approach allowed us to test the influence of the uncertainty of the surface brightness on the obtained results. The relations are also presented in Figure \ref{fig:sbcr} together with the colour index range covered by each star during the pulsation cycle.

The light and radial velocity curves were phased with $HJD_0$ set to the maximum of brightness in the $V$ band. We then fit the Akima spline polynomials \citep{10.1145/321607}, implemented in the \texttt{SciPy} Python library \citep{2020SciPy-NMeth} or template (created from ASAS or ASASSN data), to the $V$ band light curve and radial velocity curve. We found the mean radial velocity $V_{\gamma}$ by integrating the radial velocity curve, and then we calculated the following integral:

\begin{equation}\label{eq:integral}
    \Delta R'(\phi_i) = \frac{\Delta R(\phi_i)}{p} =\int _{0} ^{\phi_i} -(V_r(\phi)-V_{\gamma}) P d\phi,
\end{equation}
which gave us the displacement (in kilometres) of the physical radius $\Delta R(\phi_i)$ over the pulsation phase but divided by the projection factor $p$. We found the mean value of $\Delta R'(\phi_i)$ and subtracted it from the above integral to obtain the displacement relative to the mean radius of the star. 

From the model of the $V$ band light curve, $K_{\mathrm{S}}$ band measurements, and the reddening correction, we obtained the unreddened $(V-K_{\mathrm{S}})_0$ colour index for the phases corresponding to the phases of the $K_{\mathrm{S}}$ band measurements. From the selected SBCR, we then calculated the surface brightness $F_V$ for a given phase of $K_{\mathrm{S}}$ band measurements, and then from equation \ref{eq:sb}, we calculated the angular size $\theta$ (in miliarcseconds) of the star in a given phase. Relations from \citet{2021AA...652A..26S} and \citet{2021AA...649A.109G} are given for $K_{\mathrm{s}}$ photometry in the $2MASS$ system, but in the case of \citet{2004AA...426..297K,2004AA...428..587K}, which are in the $SAAO$ system, we transformed our $K_{\mathrm{S}}$ band photometry using equations from \citet{2007MNRAS.380.1433K}.

The angular size for a given phase $\theta(\phi)$ and the physical radius $R(\phi)$ are connected by the following geometrical formula:

\begin{equation}\label{eq:bw_0}
    \begin{aligned}
    \theta (\phi_i)= 6.7114\times10^{-9} \times 2 \omega R(\phi_i)
    \end{aligned},
\end{equation}
where $\omega$ is the parallax (in milliarcseconds) and the numerical constant comes from changing parsec to kilometres and radians to milliarcseconds (it is in fact simply the inverse of the astronomical unit). Using equation \ref{eq:integral}, we obtained the following relation:

\begin{equation}\label{eq:bw}
    \begin{aligned}
    \theta (\phi_i)= 6.7114\times10^{-9}\times2\omega(p\Delta R'(\phi_i) + \langle R \rangle)
    \end{aligned},
\end{equation}
where $\langle R \rangle$ is the mean radius of the star. By fitting a straight line to this relation, we obtained the projection factor (as a slope) and the mean angular diameter of the star (the zero point), which is used to calculate the mean radius.



\section{Results and discussion}\label{sec:results}

Figure \ref{fig:v439_oph_bw} and Figures \ref{fig:vy_pyx_bw} - \ref{fig:fm_del_bw} in the Appendix \ref{ap:fig} present the light and radial velocity curves (panels $a$,$b$,$d$), the $(V-K_{\mathrm{S}})$ colour index curve (panel $c$), the integrated radial velocity (panel $e$), and the angular diameter curve based on $K04b$ SBCR (panel f) for the analysed stars. Panel $g$ shows the linear relation between the angular diameter and the integrated radial velocity in corresponding phases rescaled using equation \ref{eq:bw} to have identical units in both axes. The slope of the fitted line is simply the $p$-factor in such a case, and the zero point is the mean angular size of the star, which is used to calculate the mean radius. In panel $f$, we also plot the integrated radial velocity curve from panel $e$, rescaled using equation \ref{eq:bw}, and the estimated values of the $p$-factor and mean radius. In the cases of VY~Pyx, SW~Tau, V971~Aql, and V439~Oph, the $V$ band light curve was approximated with Akima splines, while for BL~Her, KZ~Cen, RT~TrA, and FM~Del, we used a template created based on the ASAS or ASASSN $V$ band light curve of a given star.

\begin{figure}[h]
    \centering
    \includegraphics[width=0.5\textwidth]{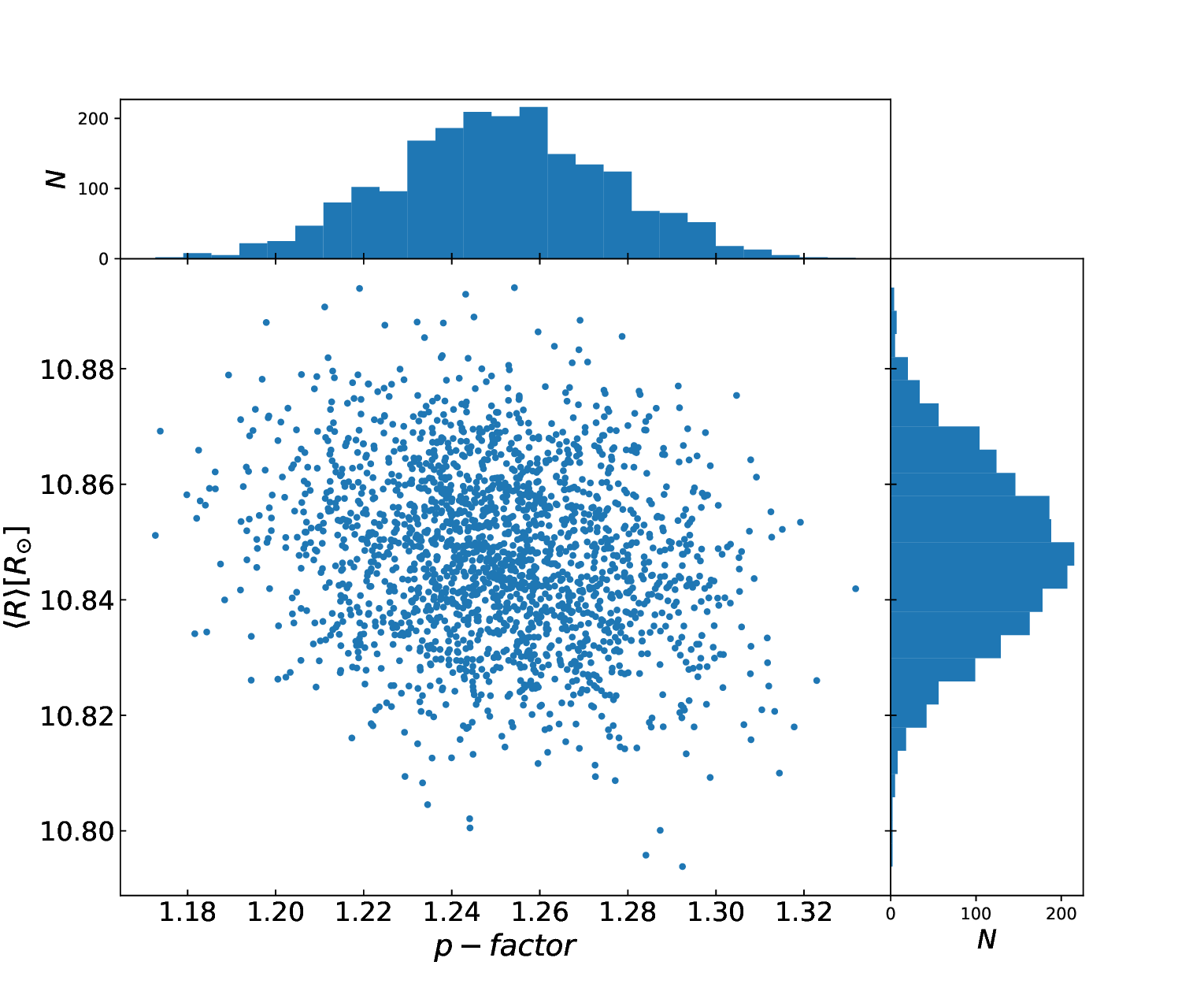}
    \caption{Scatter plot for the $p$-factor and mean radius from 2000 Monte-Carlo simulations and respective histograms for both parameters for V439 Oph.\label{fig:hist_p}}
    \end{figure}

Statistical uncertainties on the obtained values of the $p$-factors and radii were evaluated in the Monte-Carlo process. In each simulation, we drew $V$ and $K_{\mathrm{S}}$ band and radial velocity measurements from normal distributions defined by the original measurements and related errors and perform the BW analysis. We repeated this process 2000 times, and from the resulting histograms of the $p$-factor and radius, we obtained the uncertainties as the standard deviation of a Gaussian fitted to the histogram. In Figure \ref{fig:hist_p}, we show the scatter plot for the $p$-factor and mean radius obtained in Monte-Carlo simulations for V439~Oph and respective histograms for these two parameters. The Pearson correlation coefficient $r$ for the $p$-factor and mean radius for each analysed star is between -0.2 and 0.2, which means that the correlation between these two parameters is very weak. 

\begin{figure}[h]
        \centering
        \includegraphics[width=0.45\textwidth]{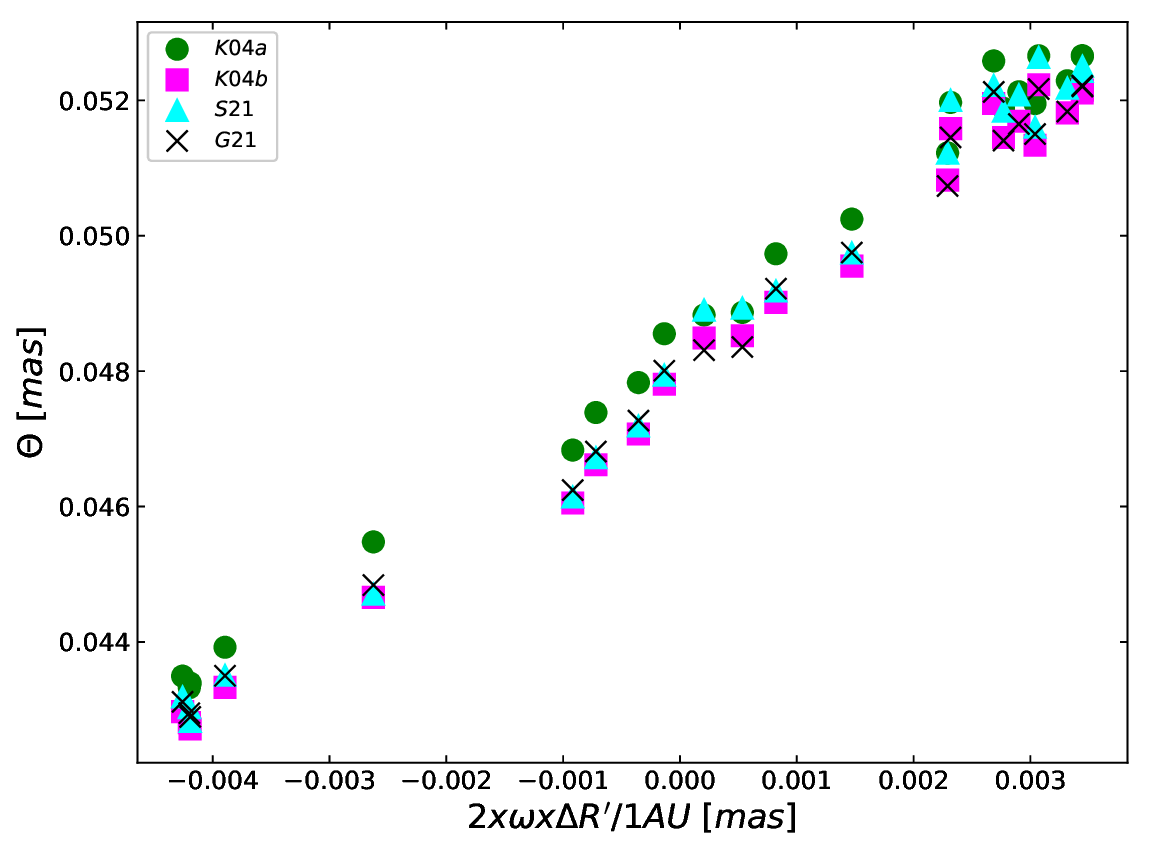}
        \caption{Relation between the angular diameter inferred from different SBCRs and the integrated radial velocity rescaled using equation \ref{eq:bw}  for V439 Oph.\label{fig:v439oph_sbcr}}
        \end{figure}

The obtained values of the $p$-factors and radii obtained using different SBCRs with statistical errors resulting from the Monte-Carlo simulations are summarised in Table \ref{tab:results}. As the relation K04b covers the whole range of the colours of all analysed Type II Cepheids and gives in most cases the smallest dispersion of the final fit, we used it as a reference relation. We used the other SBCRs to estimate the uncertainty related to the uncertainty of the surface brightness. In Figure \ref{fig:v439oph_sbcr}, we show for comparison the relation from panel $g$ of Figure \ref{fig:v439_oph_bw} for the star V439~Oph but for each SBCR used in the analysis. 

We tested the influence of the systematic errors of the $V$ and $K_{\mathrm{S}}$ band photometry, $E(B-V)$, and parallax on our results. We simply shifted a given curve or parameter by the assumed systematic value (0.02 mag for $V$ and $K_{\mathrm{S}}$ light curves, and errors from Table \ref{tab:data} for reddening and parallax), and we repeated the whole process to find the corresponding value of the $p$-factor and radius. The difference between the value obtained with the original dataset and the value obtained after applying the systematic shift gave us the uncertainty related to the uncertainty of a given parameter.


The spread of the obtained values of the $p$-factors and radii from different SBCRs was adopted as the uncertainty related to the surface brightness. The obtained uncertainties for every star are summarised in Table \ref{tab:errors}. The total errors on the $p$-factors and radii are the quadratic sum of the uncertainties detailed in this Table for each star. 

\begin{figure}[h]
            \centering
            \includegraphics[width=0.5\textwidth]{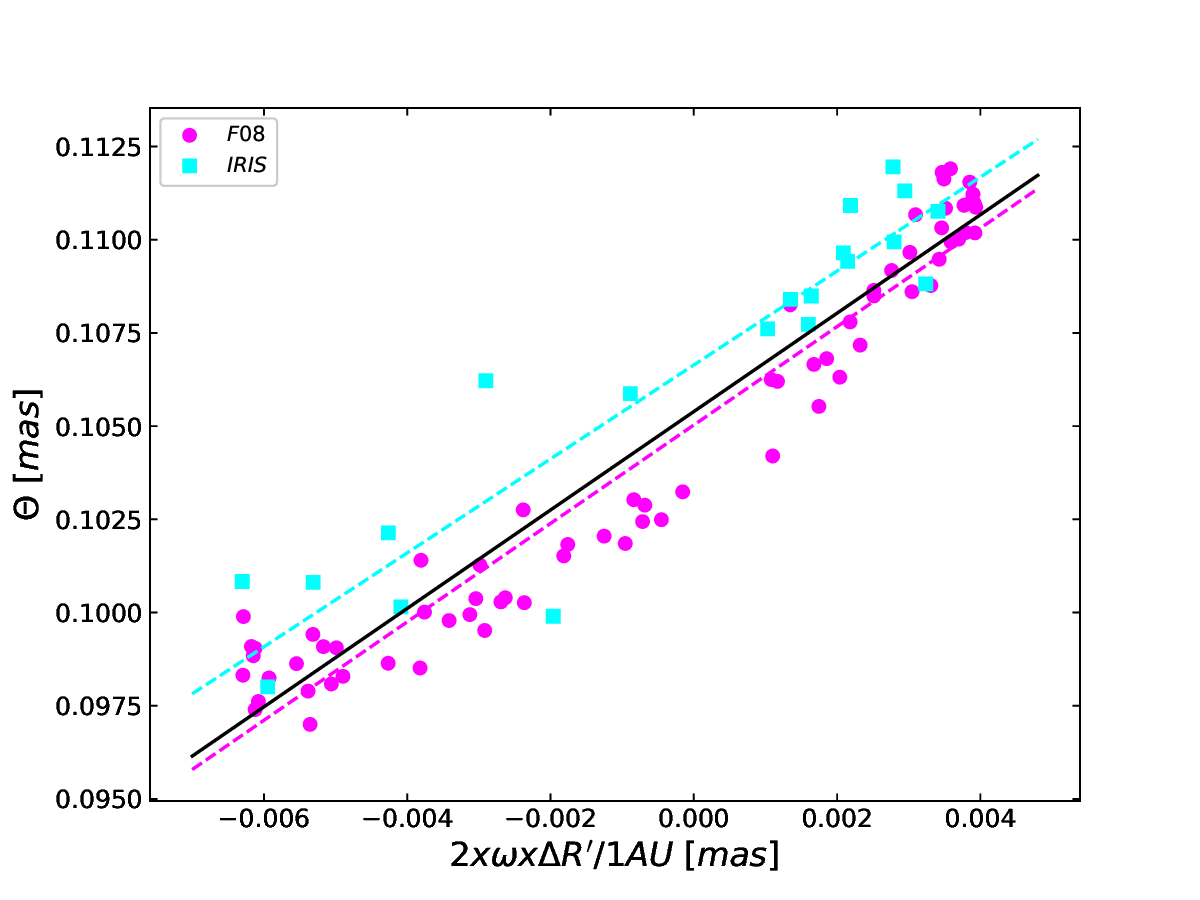}
            \caption{Baade-Wesselink fit for SW Tau exclusively using IRIS (cyan) and \citet[][F08, magenta]{2008MNRAS.386.2115F} $K_{\mathrm{S}}$ band photometry. The black line is the fit using the whole dataset.\label{fig:swtau_iris_f08}}
            \end{figure}

We note that SW~Tau is the only star for which we used the $K_{\mathrm{S}}$ band photometry from two sources, and we had to apply the period change to phase the curves. In Figure \ref{fig:swtau_iris_f08}, we plot the relation from panel $f$ of Figure \ref{fig:sw_tau_bw}, but the measurements based on different $K_{\mathrm{S}}$ band photometry sources are marked with different colours. The $p$-factor obtained based only on IRIS data amounts to 1.26$\pm$0.11, while using photometry from \citet{2008MNRAS.386.2115F} gives a value of 1.32$\pm$0.04. Both values are in agreement within the statistical error with the value obtained using the combined IRIS and \citet{2008MNRAS.386.2115F} data, which amounts to 1.32$\pm$0.04. In the case of radii, we obtained 9.29$\pm$0.03 and 9.15$\pm$0.01 when we exclusively used IRIS and \citet{2008MNRAS.386.2115F}, respectively. These values differ by $\sim$1.6\%, which is most probably caused by a small systematic difference (0.015 mag) in the $K_{\mathrm{S}}$ light curves of these two sources.

\subsection{Projection factors} \label{subsec:disc_mw}
The values of the $p$-factor for the reference $K04b$ relation mostly range between 1.21 and 1.36, and the results for different SBCRs are usually in agreement within statistical errors. One significantly outlying star, VY Pyx, is the brightest star in our sample (brighter than 6 mag in $K_{\mathrm{S}}$) and has a very low amplitude (0.3 mag in $V$ and $K_{\mathrm{S}}$). It was observed with the non-cooled neutral density filter, which increased the thermal noise and results with a low-quality $K_{\mathrm{S}}$ band light curve. These factors result in the low precision of the fit. The weighted mean value of the $p$-factors, including VY~Pyx, is 1.330$\pm$0.058, and the standard deviation amounts to 0.143. When we excluded VY~Pyx, we obtained a weighted mean of 1.314$\pm$0.022 and a standard deviation of 0.054. These mean values of the $p$-factor are in good agreement (1$\sigma$) with the values obtained for peculiar W Vir stars in \citet{2015A&A...576A..64B} and \citet{2017ApJ...842..110P}, which are 1.26$\pm$0.07 and 1.30$\pm$0.03, respectively.

From Table \ref{tab:errors}, one can see that the most significant sources of errors on $p$-factors in most cases are the statistical spread of the final BW fit, which is connected to the dispersion and coverage of light and radial velocity curves and can be minimised with a higher number of measurements, and parallax, which will be also improved in the future with the final $Gaia$ data release. As expected, the influence of reddening on the obtained results is negligible (below 1\%). The systematic shift of the $V$ band light curve gives a change of $p$-factor below 1\%, and for $K_{\mathrm{S}}$, it is at the level of 1.5\%. Errors related to the uncertainty of the surface brightness are at the level of 2-3\%.

\begin{figure}[h]
    \centering
    \includegraphics[width=0.45\textwidth]{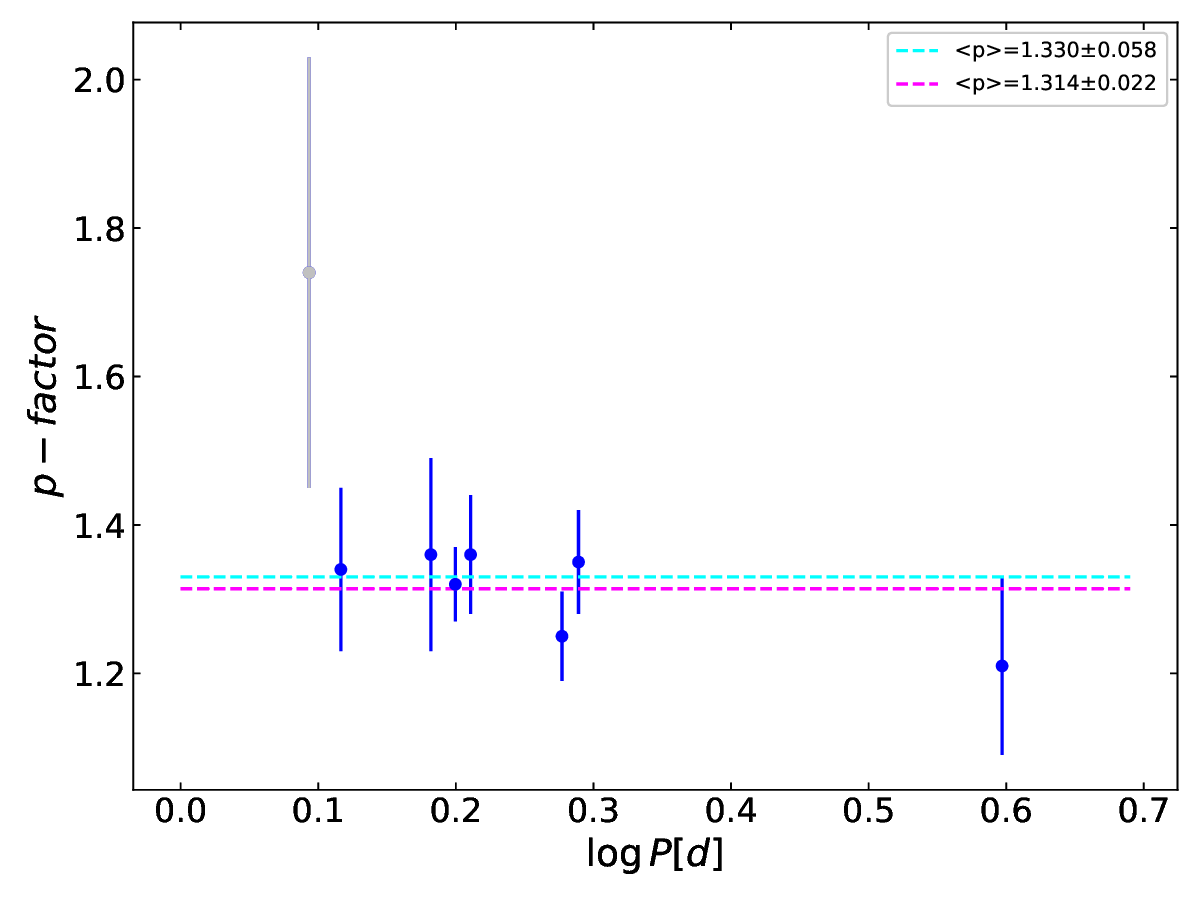}
    \caption{Period $-$ $p$-factor relation for BL Her stars. The star marked with a grey point is VY~Pyx. The dashed lines denote the weighted mean values of $p$-factors including VY~Pyx (cyan) and without this star (magenta).\label{fig:ppr}}
    \end{figure}

In Figure \ref{fig:ppr}, we plot the relation between the $p$-factors and the logarithms of periods of the analysed stars. The errors of the $p$-factors are the total errors from Table \ref{tab:errors} (every source of the error in this table contributes to the observed dispersion of the $p$-factors in this plot), and in most cases the measured values of the $p$-factors are in agreement with the mean value within 1$\sigma$. No obvious relation between the $p$-factor and the period is visible, so at this stage we can conclude that for short-period Type II Cepheids, it is sufficient to use the constant value of the $p$-factor given above.

Recently, \citet{2024zgirski} used the same method and data from identical instruments as in this work to measure the $p$-factors of nine RR Lyr stars. They obtained a mean value of 1.44$\pm$0.03, which is higher than our measurement for Type II Cepheids by $\sim$2$\sigma$.

Our mean value of the $p$-factor can be compared to those obtained recently for RR Lyrae and classical Cepheids in \citet{2024AA...684A.126B} and \citet{2021AA...656A.102T} from analysis with the \texttt{SPIPS} code, which are 1.25$\pm$0.09 and 1.24$\pm$0.14, respectively. The value obtained in this work for BL Her stars is in 1$\sigma$ agreement with these results; however, it is important to note that we used a different version of the BW method. The dispersion of the $p$-factor for BL Her stars, considering the whole analysed sample, seems to be similar to that of the RR Lyr and classical Cepheids. If we neglect the outlying star VY~Pyx, as its uncertainty is much bigger than that of the other stars and mostly statistical, we obtain a much smaller dispersion. The size of our sample is, however, too small to conclusively determine the real intrinsic dispersion of the $p$-factor for BL Her stars.

In Section \ref{sec:data}, we mentioned that the selection of the function to model the CCF has quite a significant impact on the obtained values of the $p$-factor. Indeed, $p$-factors of the analysed stars obtained using radial velocities determined from the CCF modelled with the bi-Gaussian and centroid \citep{2017AA...597A..73N} are systematically lower by $\sim$0.01 ($\sim$1\%) in the case of the centroid and 0.05 ($\sim$5\%) for the bi-Gaussian, but the spread of the $p$-factors is identical to when we used the Gaussian. Based on this, we can claim that using a Gaussian to model the CCF is a good approach for BL Her stars as long as it is used consistently for the calibration of the $p$-factor (or $p$-factor $-$ period relation) and the distance determination of the BL Her star, which requires adopting a value for its $p$-factor. In other words, different implementations of the IRSB BW technique might require different $p$-factor values for the same star.

As described in Section \ref{sec:data}, we used \citet{2021A&A...649A...4L} corrections for the zero-point offset of parallaxes. However, these corrections are still debated in the literature \citep[e.g.][]{2021AA...654A..20G,2023MNRAS.520.4154M,2023AA...672A..85C}, and many authors have suggested that they are overestimated. For comparison, we repeated the BW analysis without introducing any corrections. The resulting values of the $p$-factors are greater by 0.5-5\%, depending on the distance of the star and the size of the zero-point correction, but they agree within the error bars with the values obtained when zero-point corrections are applied. The mean value of the $p$-factor in this case is greater by 0.03, but the spread is similar in both cases, so if the observed scatter of the $p$-factor is related to the error on parallaxes, applying corrections from \citet{2021AA...649A...2L} does not improve the results.

\begin{figure*}[h]
    \centering
    \includegraphics[width=0.9\textwidth]{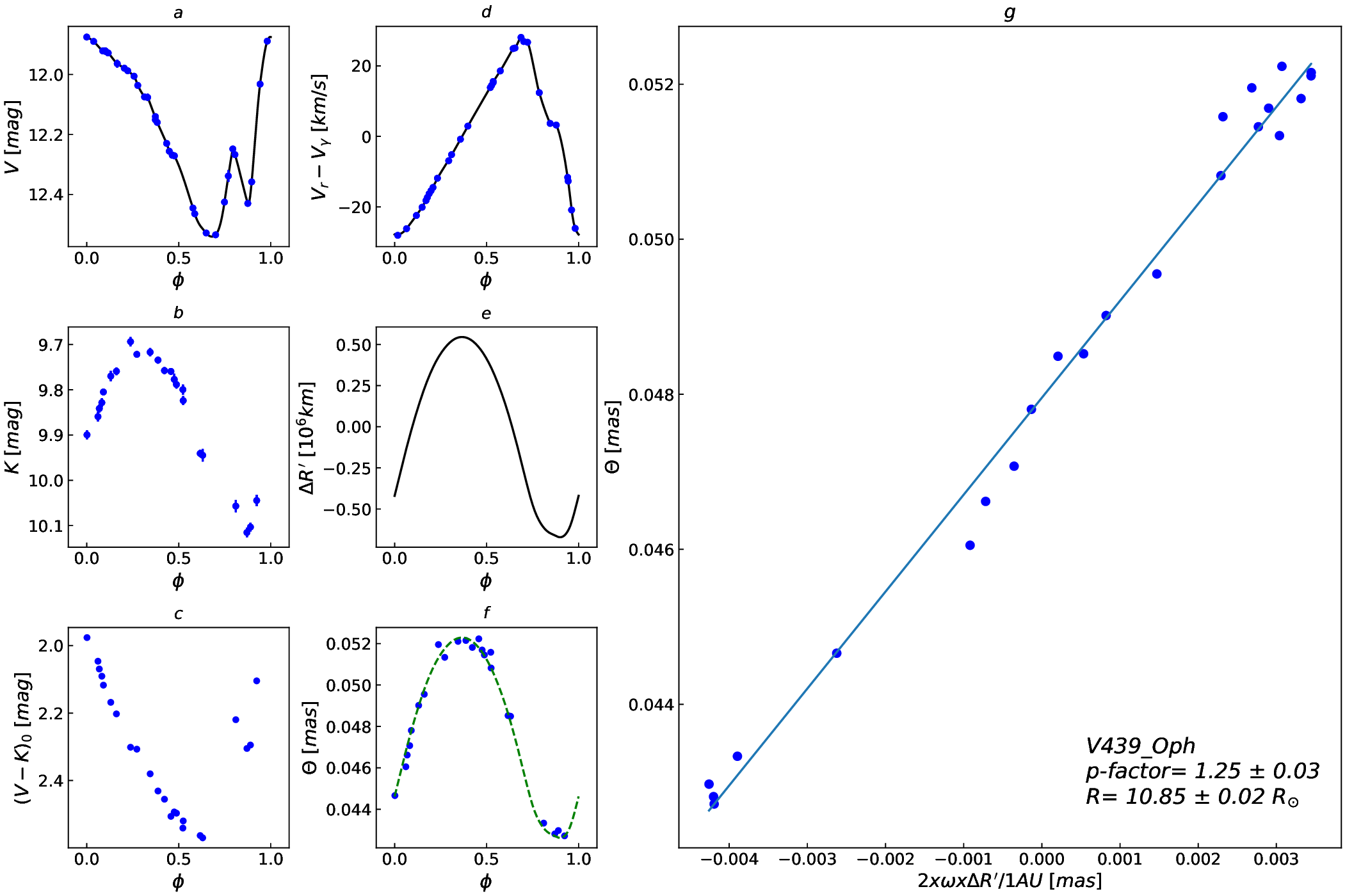}
    \caption{Baade-Wesselink analysis of V439~Oph. Panels $a$, $b$, and $d$ show our $V$, $K_{\mathrm{S}}$, and radial velocity measurements, respectively. Panel $c$ shows the unreddened colour index curve. Black lines in these plots are Akima splines fitted to the measurements. Panel $e$ shows the integrated radial velocity curve. Panel $f$ shows the angular diameter measured using the $K04b$ SBCR. The green line in this panel is the curve from panel $e$, rescaled using equation \ref{eq:bw} and measured values of the $p$-factor and the mean radius. Panel $g$ shows the relation between angular diameters from panel $f$ and corresponding values of the integrated radial velocity from panel $e$, rescaled using equation \ref{eq:bw}. The slope of this relation is the $p$-factor, and the zero point is the mean angular diameter used to calculate the mean radius of the star. \label{fig:v439_oph_bw}}
    \end{figure*}
    
\begin{table}[h]
    \caption{Projection factors and radii obtained for the analysed Type II Cepheids. The presented uncertainties are the statistical errors of the BW fit calculated in the Monte-Carlo simulations.}
    \label{tab:results}
    \centering
    \begin{tabular}{cccc}
    \hline\hline
   
    $Name$ & $SBCR$ & $p$ & $\langle R \rangle$\\
     &  & &  ($R_\odot$)  \\
    \hline
    \hline
    VY Pyx & K04a & 1.72$\pm$0.28 & 8.50$\pm$0.04\\
    & \textbf{K04b} & \textbf{1.74$\pm$0.29} & \textbf{8.38$\pm$0.05}\\
    & S21 & 1.77$\pm$0.29 & 8.40$\pm$0.05\\
    & G21 & 1.74$\pm$0.29 & 8.41$\pm$0.05\\
    \hline
    BL Her & K04a & 1.34$\pm$0.09 & 8.45$\pm$0.02 \\
    & \textbf{K04b} & \textbf{1.34$\pm$0.10} & \textbf{8.27$\pm$0.03}\\
    & S21 & 1.36$\pm$0.11 & 8.27$\pm$0.03\\
    & G21 & 1.33$\pm$0.10 & 8.31$\pm$0.03\\
    \hline
    KZ Cen & K04a & 1.28$\pm$0.08 & 9.42$\pm$0.03\\
    & \textbf{K04b} & \textbf{1.35$\pm$0.08} & \textbf{9.25$\pm$0.03}\\
    & S21 & 1.42$\pm$0.09 & 9.26$\pm$0.03\\
    & G21 & 1.35$\pm$0.08 & 9.29$\pm$0.03\\
    \hline
    SW Tau & K04a & 1.30$\pm$0.04 & 9.38$\pm$0.01\\
    & \textbf{K04b} & \textbf{1.32$\pm$0.04} & \textbf{9.18$\pm$0.01}\\
    & S21 & 1.34$\pm$0.05 & 9.17$\pm$0.02\\
    & G21 & 1.31$\pm$0.04 & 9.23$\pm$0.01\\
    \hline
    V971 Aql & K04a & 1.35$\pm$0.03 & 9.73$\pm$0.01\\
    & \textbf{K04b} & \textbf{1.36$\pm$0.03} & \textbf{9.58$\pm$0.01}\\
    & S21 & 1.39$\pm$0.03 & 9.61$\pm$0.01\\
    & G21 & 1.36$\pm$0.03 & 9.62$\pm$0.01\\
    \hline
    V439 Oph & K04a & 1.23$\pm$0.03 & 10.98$\pm$0.02\\
    & \textbf{K04b} & \textbf{1.25$\pm$0.03} & \textbf{10.85$\pm$0.02}\\
    & S21 & 1.28$\pm$0.03 & 10.91$\pm$0.02\\
    & G21 & 1.23$\pm$0.03 & 10.87$\pm$0.02\\
    \hline
    RT TrA & K04a & 1.31$\pm$0.04 & 10.30$\pm$0.02\\
    & \textbf{K04b} & \textbf{1.35$\pm$0.05} & \textbf{10.13$\pm$0.03}\\
    & S21 & 1.39$\pm$0.06 & 10.16$\pm$0.03\\
    & G21 & 1.35$\pm$0.05 & 10.17$\pm$0.03\\
    \hline
    FM Del & K04a & 1.19$\pm$0.06 & 15.28$\pm$0.05\\
    & \textbf{K04b} & \textbf{1.21$\pm$0.07} & \textbf{15.11$\pm$0.05}\\
    & S21 & 1.24$\pm$0.08 & 15.20$\pm$0.06\\
    & G21 & 1.19$\pm$0.06 & 15.12$\pm$0.05\\
    \hline
    \end{tabular}
    \end{table}

    \begin{table*}
        \caption{Summary of the errors obtained for $p$-factors and radii for each star.}
        \label{tab:errors}
        \centering
        \begin{tabular}{c|ccccccc|ccccccc}
        \hline\hline
       
        $Name$ & $\sigma_{stat}$ & $\sigma_{SBCR}$ &  $\sigma_{V}$ & $\sigma_{K_{\mathrm{S}}}$ & $\sigma_{E(B-V)}$ & $\sigma_{\omega}$  & \textbf{$\sigma_{total}$} & $\sigma_{stat}$ & $\sigma_{SBCR}$ & $\sigma_{V}$ & $\sigma_{K_{\mathrm{S}}}$ & $\sigma_{E(B-V)}$ & $\sigma_{\omega}$  & \textbf{$\sigma_{total}$}\\
        \hline
        & \multicolumn{7}{c}{$p$-factors} & \multicolumn{7}{c}{$\langle R \rangle$ ($R_{\odot}$)}\\
        \hline
        VY Pyx & 0.29 & 0.02	& 0.01	& 0.03	& 0.01	& 0.01 & \textbf{0.29} & 0.05 & 0.05	& 0.03 & 0.11 & 0.05 & 0.04 & \textbf{0.15}\\
        BL Her & 0.10 & 0.01	& 0.01	& 0.02	& 0.01	& 0.03 & \textbf{0.11} & 0.03 & 0.09	& 0.03 & 0.13 & 0.05 & 0.18 & \textbf{0.25}\\
        KZ Cen & 0.08 & 0.06	& 0.01	& 0.02	& 0.01	& 0.08 & \textbf{0.13} & 0.03 & 0.09	& 0.03 & 0.15 & 0.06 & 0.52 & \textbf{0.55}\\
        SW Tau & 0.04 & 0.02	& 0.01	& 0.02	& 0.01	& 0.02 & \textbf{0.05} & 0.03 & 0.10	& 0.04 & 0.15 & 0.06 & 0.17 & \textbf{0.26}\\
        V971 Aql & 0.03 & 0.02	& 0.01	& 0.02	& 0.01	& 0.07 & \textbf{0.08} & 0.01 & 0.07	& 0.03 & 0.12 & 0.06 & 0.48 & \textbf{0.50}\\
        V439 Oph & 0.03 & 0.02	& 0.01	& 0.02	& 0.01	& 0.04 & \textbf{0.06} & 0.02 & 0.06	& 0.04 & 0.17 & 0.07 & 0.39 & \textbf{0.44}\\
        RT TrA & 0.05 & 0.03	& 0.01	& 0.02	& 0.01	& 0.02 & \textbf{0.07} & 0.03 & 0.07	& 0.04 & 0.13 & 0.06 & 0.17 & \textbf{0.24}\\
        FM Del & 0.07 & 0.04	& 0.01	& 0.02	& 0.01	& 0.08 & \textbf{0.12} & 0.05 & 0.09	& 0.04 & 0.20 & 0.10 & 1.04 & \textbf{1.07}\\
        
        \hline
        \end{tabular}
        \tablefoot{ The total errors $\sigma_{total}$ are the quadratic sum of the statistical errors of the BW fit $\sigma_{stat}$, errors related to the uncertainty of the surface brightness $\sigma_{SBCR}$, errors related to the systematic shift of the $V$ band light curve $\sigma_{V}$, errors related to the systematic shift of the $K_{\mathrm{S}}$ band light curve $\sigma_{K_{\mathrm{S}}}$, errors related to the $E(B-V)$ uncertainty $\sigma_{E(B-V)}$, and errors related to the uncertainty of the parllax $\sigma_{\omega}$.}
        \end{table*}

\subsection{Radii} \label{subsec:disc_lmc}
The dominating sources of uncertainty on radii are the $K_{\mathrm{S}}$ band light curve systematic uncertainty ($\sim$2\%) and the parallax, depending on the distance of the star. In the case of VY~Pyx, the closest star in our sample, we obtained a total uncertainty of 2\% on the measured radius. For FM~Del, the most distant star in our sample, the total uncertainty is 7\%. The uncertainty related to the $V$ band systematic shift and reddening is negligible, and the uncertainty related to the surface brightness is at a level of 1\%.

Radii obtained from the parallaxes without applying \citet{2021A&A...649A...4L} zero-point offset corrections are greater by 0.5-8\% (3\% on average), depending on the distance of the star and the size of correction but are in agreement within the error bars with values obtained when corrections are applied. The spread of the measurements around the fitted period-radius relation is similar to when we use \citet{2021A&A...649A...4L} corrections.

\begin{figure*}[h]
    \centering
    \includegraphics[width=0.8\textwidth]{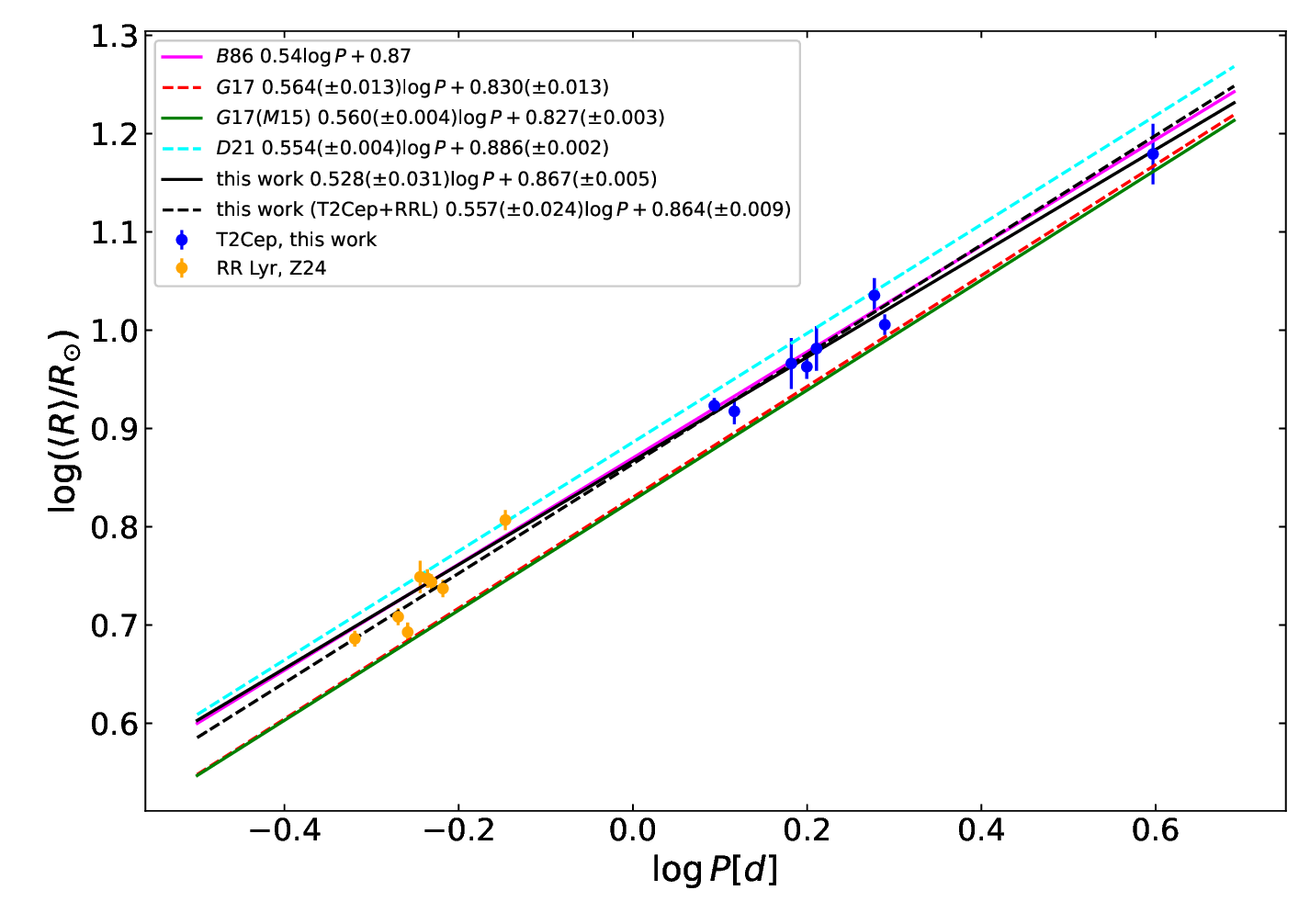}
    \caption{Period-radius relation for BL Her type stars and the common relation for BL Her and RR Lyr stars based on radii obtained in this work (blue points) and in \citep[][orange points]{2024zgirski}. For comparison, we have plotted the relations from the literature: B86 is the empirical relation \citet{1986AA...159..261B}; G17 and G17(M15) are empirical and theoretical relations from \citet{2017AA...603A..70G}, respectively; and D21 is the theoretical relation from \citet{2021MNRAS.501..875D}.\label{fig:prr}}
    \end{figure*}

Figure \ref{fig:prr} shows the relation between logarithms of radii and periods of the analysed stars. The errors of the radii in this plot are the total errors from Table \ref{tab:errors}. As expected, stars with longer periods are bigger, and the relation is linear and quite tight. The line fitted to this relation is  
\begin{equation}\label{eq:pr}
    \begin{aligned}
    \log \langle R \rangle = 0.528 (\pm 0.031) \log P + 0.867 (\pm 0.005) 
    \end{aligned}.
\end{equation}
We observed in this figure that two stars with very similar periods (namely, V439 Oph and RT TrA) differ quite significantly in radius (about 2$\sigma$). It is rather real (not related to any errors), as the bigger star (V439 Oph) has a lower effective temperature (redder colour index what is visible in Figure \ref{fig:sbcr}), and thus it is located closer to the red edge of the instability strip.

For comparison, we plotted the period-radius relations from the literature. \citet{1986AA...159..261B} presented an empirical relation for Type II Cepheids based on radii measured in their BW analysis using optical photometry (which is more sensitive to reddening) and assuming a value of 1.36 for the $p$-factor, and their relation is in good agreement with our result. Another empirical relation was published by \citet{2017AA...603A..70G} based on the temperatures obtained from the spectral energy distribution fitting and luminosities of Type II Cepheids in the Magellanic Clouds. This relation was derived for short- (BL Her stars) and long-period Type II Cepheids (W Virginis and RV Tau stars), and it predicts radii smaller by $\sim$8\%, but this might be the result of the possibly different mean metallicity of Type II Cepheid populations in the Magellanic Clouds and solar neighbourhood \citep[the latter is almost solar,][]{2022ApJ...927...89W}. The slope of the \citet{2017AA...603A..70G} relation is steeper than ours, but it is in agreement within 2$\sigma$. We also plotted two theoretical relations. \citet{2017AA...603A..70G} published a theoretical relation that is in fact the relation from \citet{2015ApJ...808...50M} based on the non-linear, convective hydrodynamical models of RR Lyr, which according to \citet{2017AA...603A..70G} is relevant for Type II Cepheids. We calculated the zero point of this relation for the solar metallicity. This relation is in a good agreement with the empirical relation from \citet{2017AA...603A..70G} and also predicts radii smaller by $\sim$8\% compared to this work. The second theoretical relation is from \citet{2021MNRAS.501..875D} and was obtained based on non-linear convective models calculated with the \texttt{RSP-MESA} code \citep{2008AcA....58..193S,2019ApJS..243...10P}. They considered different input parameters, and we plotted the relation for the most complex convection parameter set. This relation gives a radii greater by $\sim$5\% compared to our analysis, and the slope is again steeper than that we obtained.

Moreover, we plot in the same figure, the radii of RR Lyr measured with the data collected with identical instruments and analysed with the same BW method by \citet{2024zgirski}. Some authors \citep[e.g.][]{1986AA...159..261B,2015ApJ...808...50M} have noticed that these two classes of stars can obey single period-radius (and period-luminosity) relations. As can be seen in Figure \ref{fig:prr}, BL~Her and RR~Lyr stars can indeed be interpreted as having a common period-radius relation. We fit a common line to BL Her and RR Lyr stars, and our result was
\begin{equation}\label{eq:pr}
    \begin{aligned}
    \log \langle R \rangle = 0.557 (\pm 0.024) \log P + 0.864 (\pm 0.009),
    \end{aligned}
\end{equation}
with the slope in perfect agreement with \citep{2017AA...603A..70G} empirical and theoretical relations and the \citep{2021MNRAS.501..875D} theoretical relation. The exact form of the common relation, including metallicity dependence, requires more measurements of the radii and individual metallicities of both Type II Cepheids and RR Lyr stars.

\section{Summary}\label{sec:summ}
We used the homogeneous optical and near-infrared photometry and spectra collected within a short span of time and the SBCR variant of the BW method to determine projection factors and radii of eight nearby short-period Type II Cepheids. The obtained values of the $p$-factors for good quality data (seven stars) are between 1.21 and 1.36, with a typical uncertainty of $\sim$7\%. The mean value amounts to 1.330$\pm$0.058 (4\% uncertainty). The most significant sources of uncertainty are the dispersion of photometry and radial velocity curves (statistical uncertainty of the final BW fit) and $Gaia$ parallaxes. The light and radial velocity curves can be improved by collecting more data, and $Gaia$ parallaxes will be more accurate in the future data release. We did not observe any significant dependence of the $p$-factor on the period of stars.

In the case of radii, the dominant sources of uncertainty are the systematic uncertainty of the $K_{\mathrm{S}}$ band light curve and parallaxes. We derived the period-radius relations for BL Her type stars and the common relation for BL Her and RR Lyr stars \citep[using radii of RR Lyr from][]{2024zgirski}. The obtained relation for BL Her stars is in good agreement with the previous empirical relation from \citet{1986AA...159..261B} obtained from BW analysis of Galactic Type II Cepheids. The theoretical relation from \citep{2017AA...603A..70G} predicts radii systematically smaller by $\sim$8\%, while the relation from \citet{2021MNRAS.501..875D} predicts radii bigger by $\sim$5\%. The common relation for BL Her and RR Lyr stars obtained in this work has a slope that is in very good agreement with theoretical relations.

In the future, we plan to collect data and perform BW analysis of more nearby BL Her type stars as well as of long-period Type II Cepheids (i.e. W Vir type stars). This will allow us to calibrate more precisely both the $p$-factors and the period-radius relation.

\begin{center}
ACKNOWLEDGMENTS
\end{center}

P.W. gratefully acknowledges financial support from the Polish National Science Center grant PRELUDIUM 2018/31/N/ST9/02742. The research leading to these results has received funding from the European Research Council (ERC) under the European Union’s Horizon 2020 research and innovation programme (grant agreements No 695099 and No 951549). Support from DIR/2024/WK/02 grant of the Polish Ministry of Science and Higher Education and the Polish National Science Center grants MAESTRO 2017/26/A/ST9/00446 and BEETHOVEN 2018/31/G/ST9/03050 and Polish-French Marie Skłodowska-Curie and Pierre Curie Science Prize awarded by the Foundation for Polish Science is also acknowledged. W.G. and G.P. gratefully acknowledge financial support for this work from the BASAL Centro de Astrofisica y Tecnologias Afines (CATA) AFB-170002. N.N. acknowledges the support of the French Agence Nationale de la Recherche (ANR) under grant ANR-23-CE31-0009-01 (Unlock-pfactor)

Based on data collected under the ESO/CAMK PAN – USB agreement at the ESO Paranal Observatory. We thank our colleagues involved in the process of gathering data in OCM.

Based on observations collected at the European Organisation for Astronomical Research in the Southern Hemisphere under ESO programmes 099.D-0380(A), 0100.D-0339(B),0100.D-0273(A),0102.D-0281(A), 105.20L8.002, 106.20Z1.001, 106.20Z1.002, 106.21T1.001, 108.22JX.001, 111.24YL.001 and CNTAC programmes CN2016B-150, CN2017A-121, CN2017B-43, CN2018A-40, CN2019B-64, CN2020B-42, CN2020B-69. We are greatly indebted to the staff at the ESO La Silla and Paranal Observatories for excellent support during the many visitor mode runs and for performing the observations in the service mode.

This work has made use of data from the European Space Agency (ESA) mission
{\it Gaia} (\url{https://www.cosmos.esa.int/gaia}), processed by the {\it Gaia}
Data Processing and Analysis Consortium (DPAC,
\url{https://www.cosmos.esa.int/web/gaia/dpac/consortium}). Funding for the DPAC
has been provided by national institutions, in particular the institutions
participating in the {\it Gaia} Multilateral Agreement.

This publication makes use of data products from the Two Micron All Sky Survey \citep{2006AJ....131.1163S}, which is a joint project of the University of Massachusetts and the Infrared Processing and Analysis Center/California Institute of Technology, funded by the National Aeronautics and Space Administration and the National Science Foundation. This research has made use of the SIMBAD database, operated at CDS, Strasbourg, France \citep{2000A&AS..143....9W}. We acknowledge with thanks the variable star observations from the AAVSO International Database contributed by observers worldwide and used in this research.

Software used in this work: \texttt{gaiadr3$\_$zeropoint} \citep{2021A&A...649A...4L}, \texttt{Astropy7} \citep{2013A&A...558A..33A,2018AJ....156..123A}, \texttt{IRAF} \citep{1986SPIE..627..733T,1993ASPC...52..173T}, \texttt{Sextractor} \citep{1996A&AS..117..393B}, \texttt{SCAMP} \citep{2006ASPC..351..112B}, \texttt{SWARP} \citep{2010ascl.soft10068B}, \texttt{DAOPHOT} \citep{1987PASP...99..191S}, \texttt{NumPy} \citep{2011CSE....13b..22V,2020Natur.585..357H}, \texttt{SciPy} \citep{2020SciPy-NMeth}, \texttt{Matplotlib} \citep{2007CSE.....9...90H}. The custom software with GUI used for the Baade-Wesselink IRSB analysis is available on Github: \url{https://github.com/araucaria-project/balwan.git}.

\bibliographystyle{aa}
\bibliography{t2cep}

\begin{appendix}
    \section{Figures with the BW analysis}\label{ap:fig}
    
\begin{figure*}[h]
    \centering
    \includegraphics[width=0.9\textwidth]{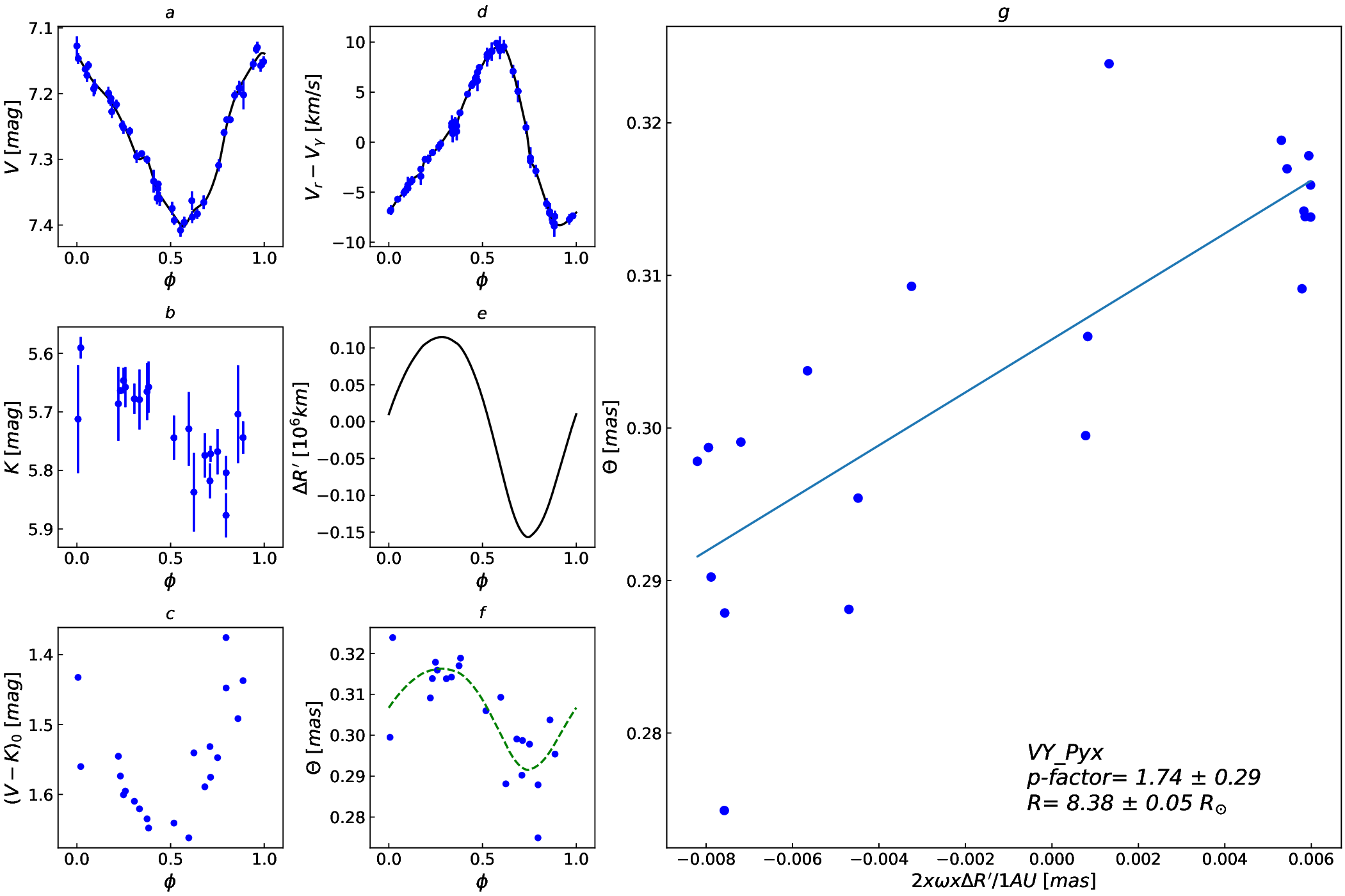}
    \caption{Baade-Wesselink analysis of VY Pyx. For a description of the panels, see Figure \ref{fig:vy_pyx_bw}.\label{fig:vy_pyx_bw}}
    \end{figure*}

    \begin{figure*}[h]
        \centering
        \includegraphics[width=0.9\textwidth]{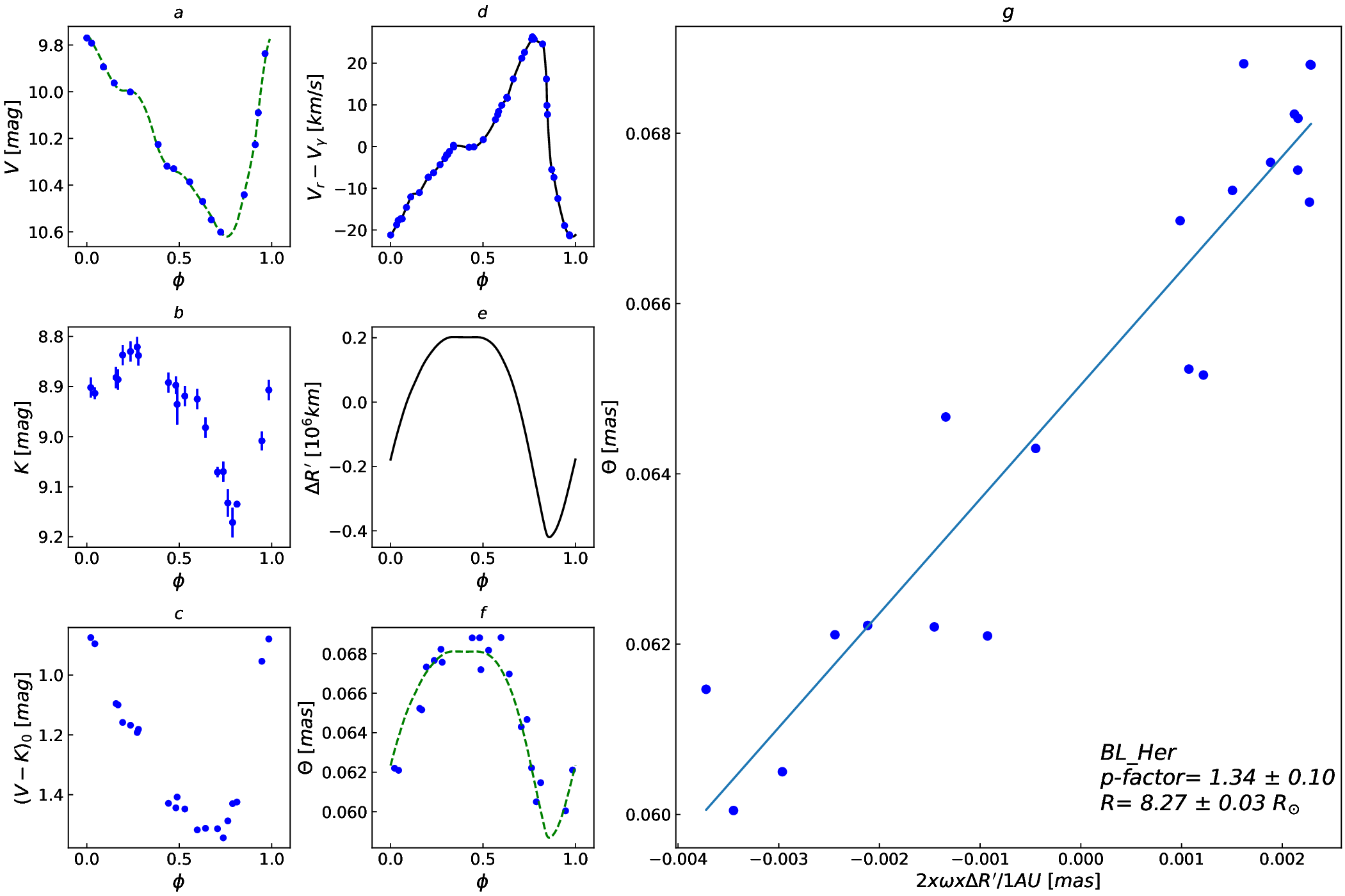}
        \caption{Baade-Wesselink analysis of BL~Her. For a description of the panels, see Figure \ref{fig:vy_pyx_bw}.\label{fig:bl_her_bw}}
        \end{figure*}

        \clearpage

    \begin{figure*}[h]
    \centering
    \includegraphics[width=0.9\textwidth]{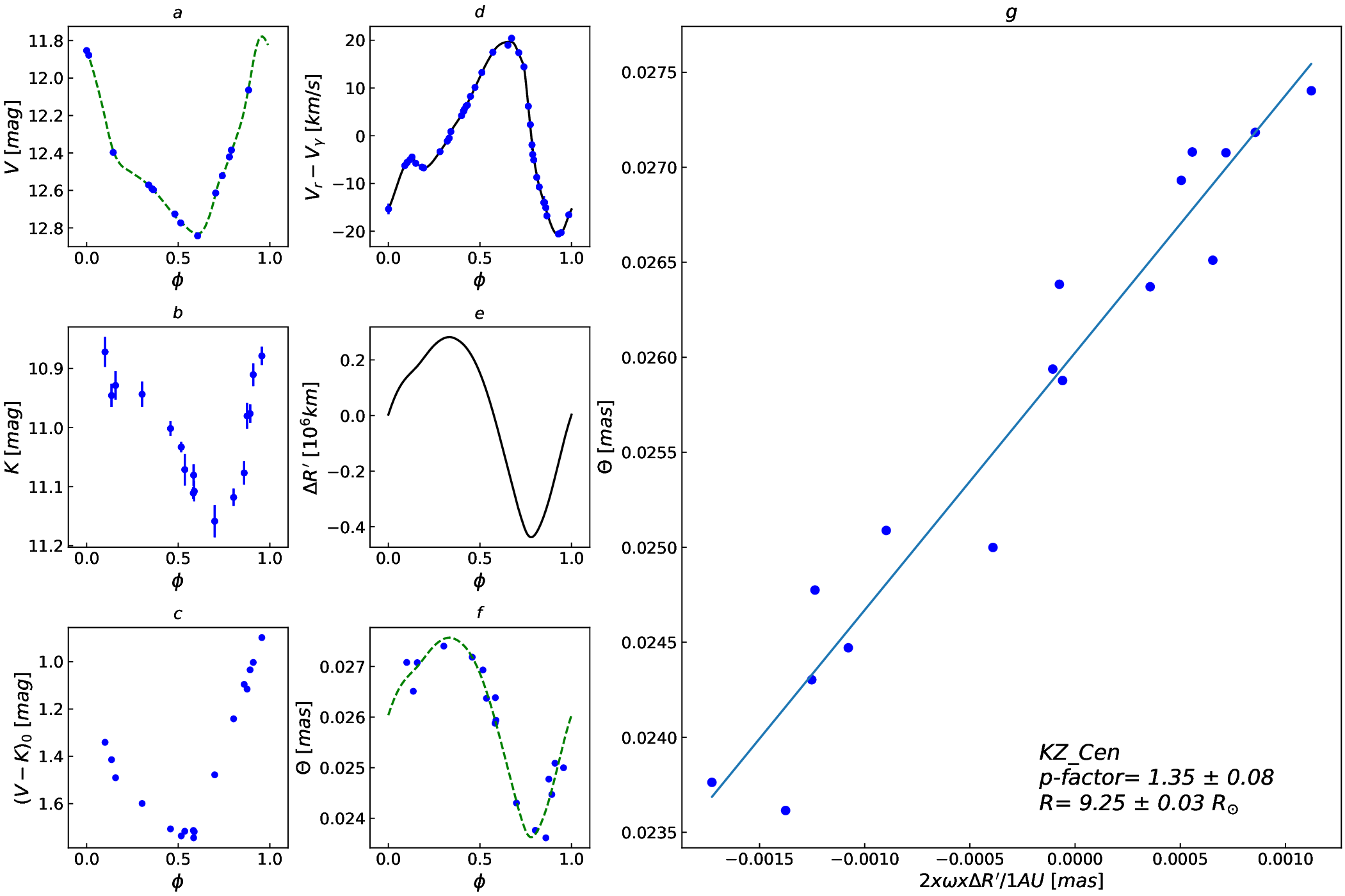}
    \caption{Baade-Wesselink analysis of KZ Cen. For a description of the panels, see Figure \ref{fig:vy_pyx_bw}.\label{fig:kz_cen_bw}}
    \end{figure*}

    \begin{figure*}[h]
        \centering
        \includegraphics[width=0.9\textwidth]{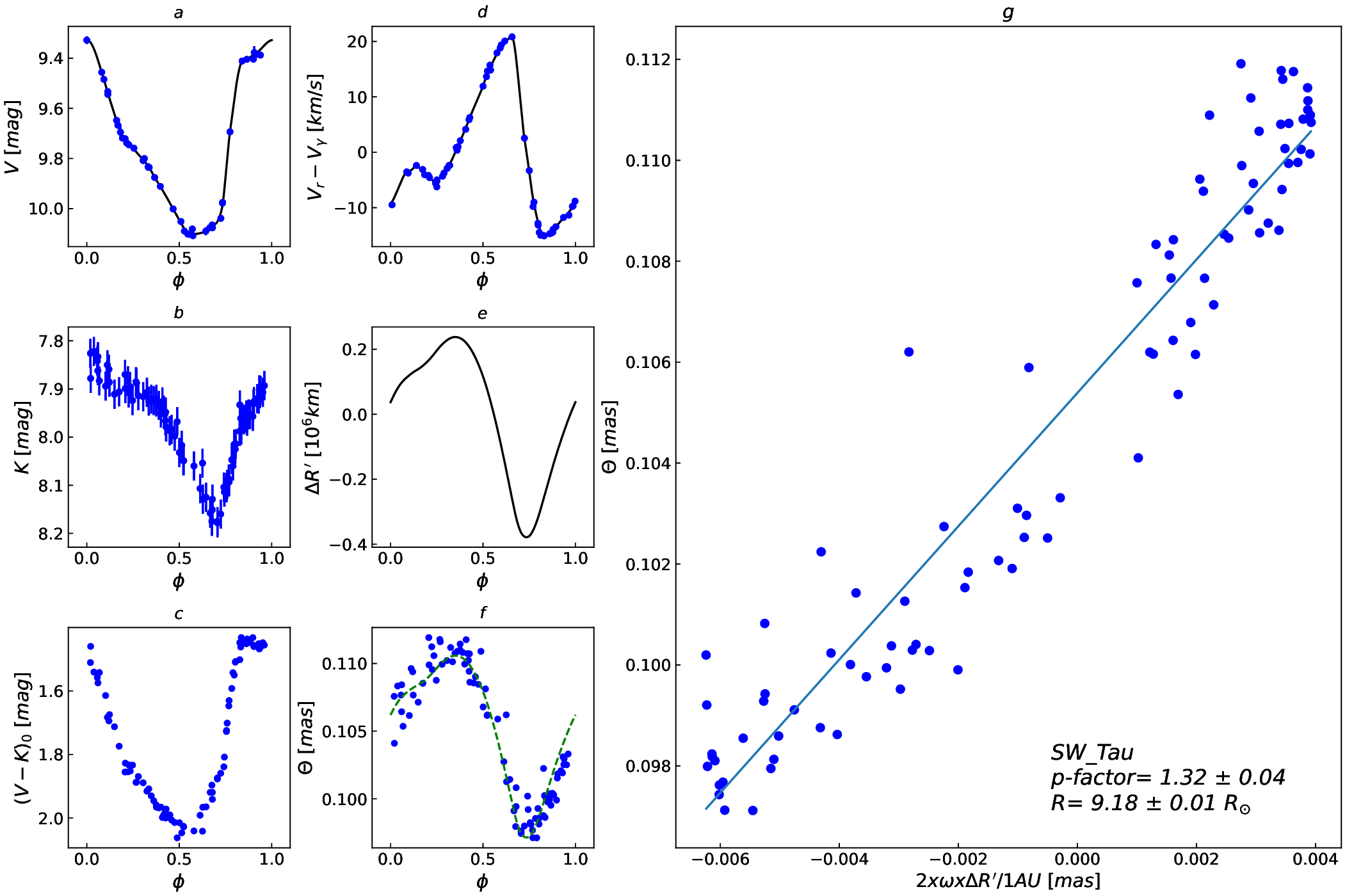}
        \caption{Baade-Wesselink analysis of SW Tau. For a description of the panels, see Figure \ref{fig:vy_pyx_bw}.\label{fig:sw_tau_bw}}
        \end{figure*}  

        \clearpage
        
    \begin{figure*}[h]
    \centering
    \includegraphics[width=0.9\textwidth]{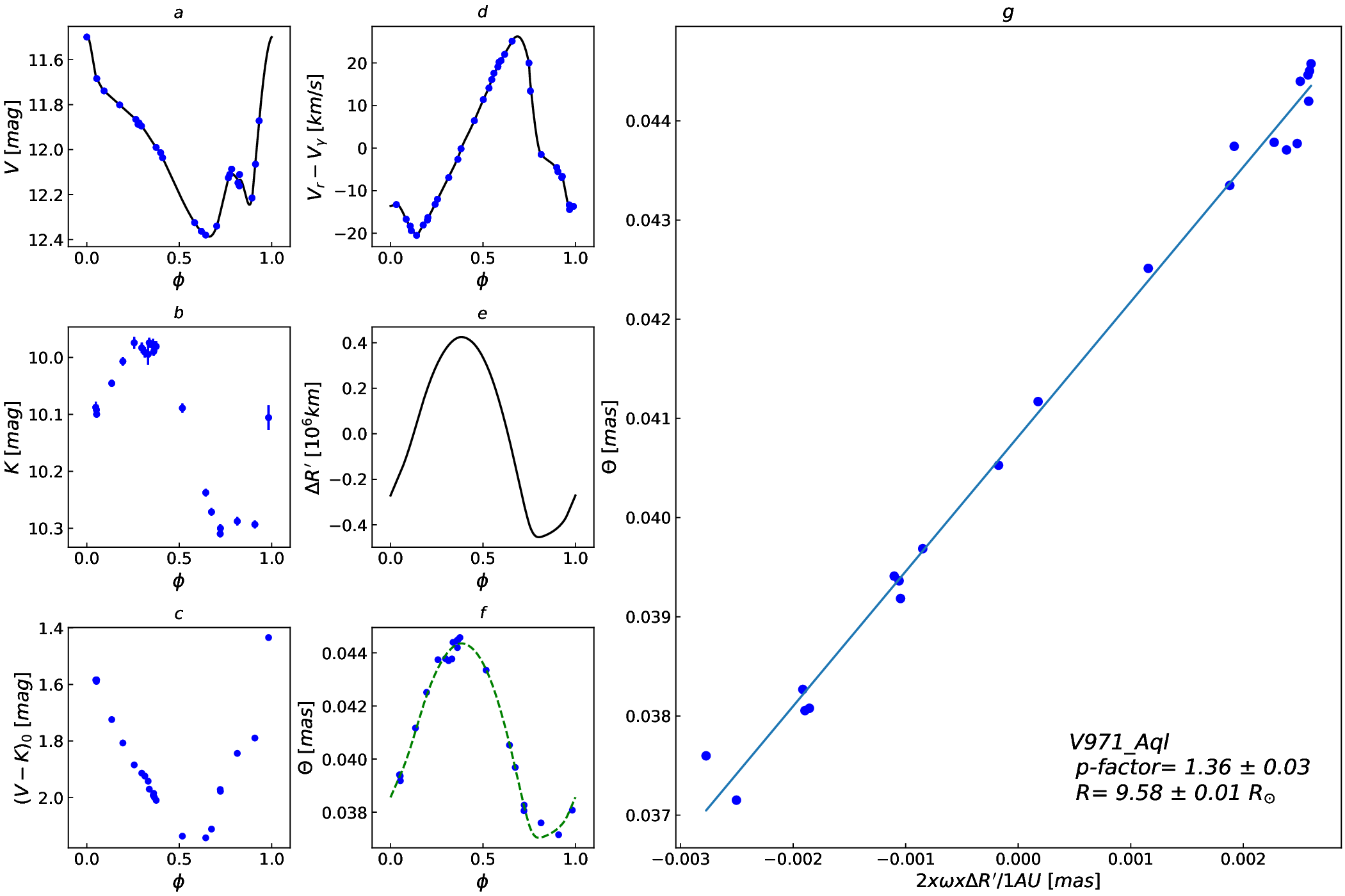}
    \caption{Baade-Wesselink analysis of V971 Aql. For a description of the panels, see Figure \ref{fig:vy_pyx_bw}.\label{fig:v971_aql_bw}}
    \end{figure*}
   
    \begin{figure*}[h]
    \centering
    \includegraphics[width=0.9\textwidth]{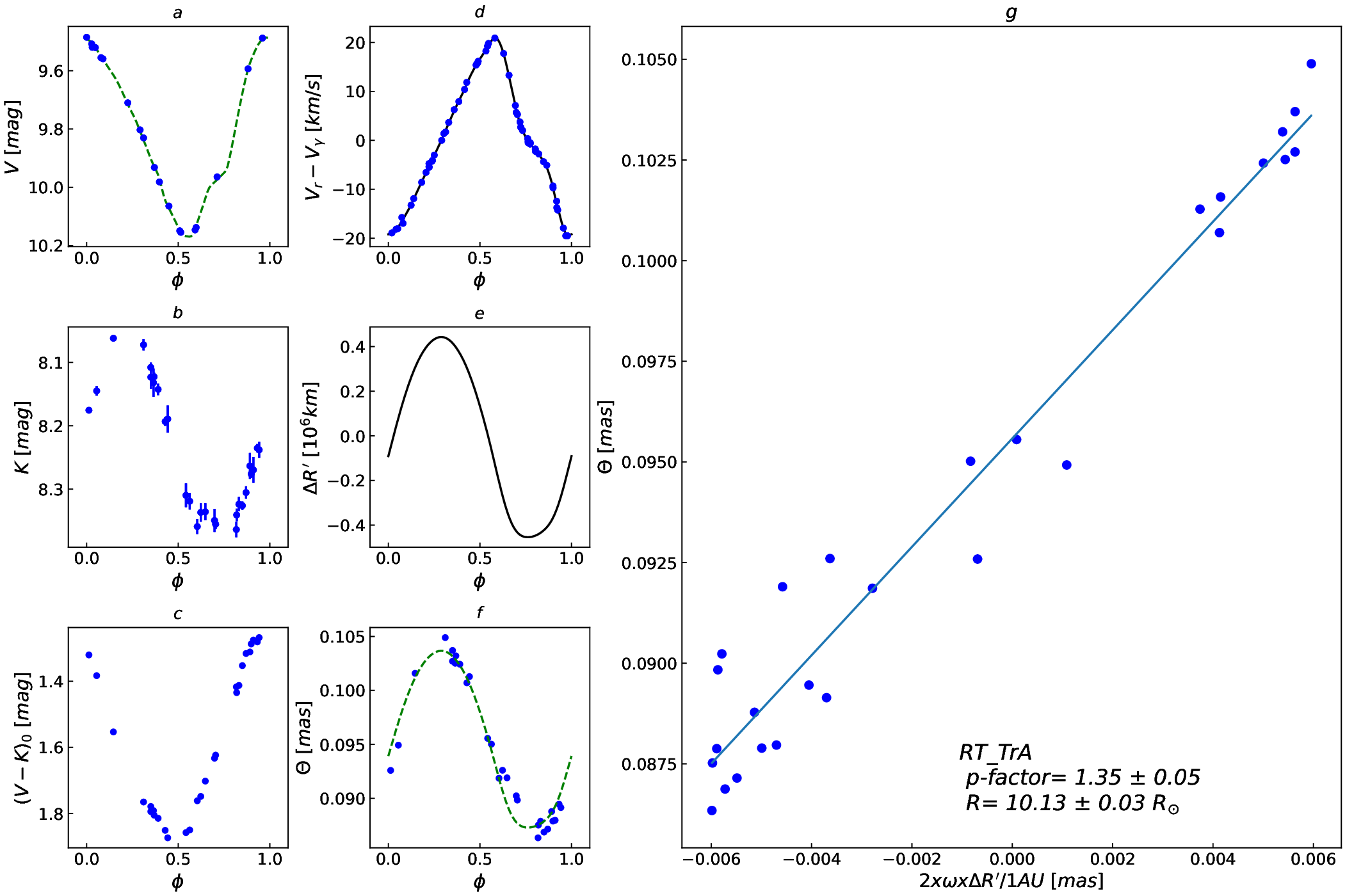}
    \caption{Baade-Wesselink analysis of RT TrA. For a description of the panels, see Figure \ref{fig:vy_pyx_bw}.\label{fig:rt_tra_bw}}
    \end{figure*}

    \clearpage
\begin{figure*}[h]
\centering
\includegraphics[width=0.9\textwidth]{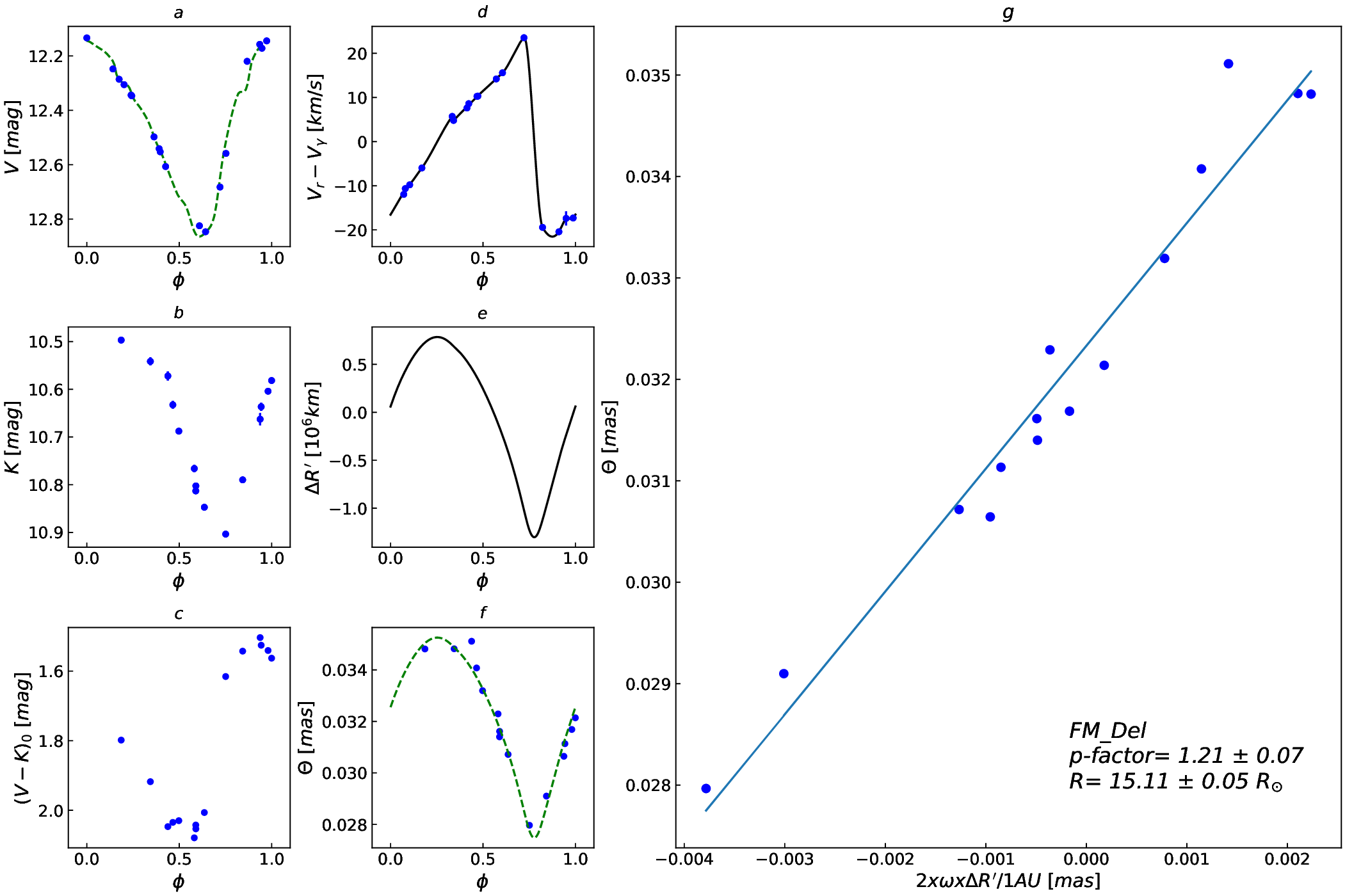}
\caption{Baade-Wesselink analysis of FM Del. For a description of the panels, see Figure \ref{fig:vy_pyx_bw}.\label{fig:fm_del_bw}}
\end{figure*}

\end{appendix}



\end{document}